\documentclass[12pt]{iopart}
\usepackage{epsfig,epsf,graphics,slashed,amssymb,subfigure}

\begin{document}

\title[Quantized vortices and superflow in arbitrary dimensions]
{Quantized vortices and superflow in arbitrary dimensions:
Structure, energetics and dynamics}

\author{Paul M.~Goldbart and Florin Bora}
\address{Department of Physics,
University of Illinois at Urbana-Champaign,\\
1110 West Green Street, Urbana, Illinois 61801-3080, U.S.A.}
\ead{goldbart@illinois.edu}

  \date{30 November 2008}

\begin{abstract}
The structure and energetics of superflow around quantized vortices, and the motion inherited by these vortices from this superflow, are explored in the general setting of the superfluidity of helium-four in arbitrary dimensions.  The vortices may be idealized as objects of co-dimension two, such as one-dimensional loops and two-dimensional closed surfaces, respectively, in the cases of three- and four-dimensional superfluidity.  By using the analogy between vorticial superflow and Amp\`ere-Maxwell magnetostatics, the equilibrium superflow containing any specified collection of vortices is constructed.  The energy of the superflow is found to take on a simple form for vortices that are smooth and asymptotically large, compared with the vortex core size.  The motion of vortices is analyzed in general, as well as for the special cases of hyper-spherical and weakly distorted hyper-planar vortices.  In all dimensions, vortex motion reflects vortex geometry.  In dimension four and higher, this includes not only extrinsic but also intrinsic aspects of the vortex shape, which enter via the first and second fundamental forms of classical geometry.  For hyper-spherical vortices, which generalize the vortex rings of three dimensional superfluidity, the energy-momentum relation is determined.  Simple scaling arguments recover the essential features of these results, up to numerical and logarithmic factors.
\end{abstract}



\maketitle

\def\deflambda{\bar{d}}
\def\deldef{\Delta}
\def\hodge{\star}
\def\perm{\hat\pi}
\def\Dsl{\slashed{D}}
\def\text{\rm}
\def\vel{V}
\def\velg{{\cal V}}
\def\velv{U}
\def\source{J}
\def\sourcedual{\hodge\,{\source}}
\def\pos{x}
\def\bpos{\bar{\pos}}
\def\dfc{R}
\def\dfcsc{\widehat{\dfc}}
\def\inin{a}
\def\gp{A}
\def\newpart{\partial}
\def\dirac{\delta}
\def\diracbar{\hat{\dirac}}
\def\bd{\bar{d}}
\def\vol{\Omega}
\def\pvol{\widehat{\vol}}
\def\mdc{\rho}
\def\dbar{\hat{d}}
\def\Phase{\Phi}
\def\opf{\Psi}
\def\sfop{\opf}
\def\amp{{\cal A}}
\def\vcs{\xi}
\def\vsk{{\kappa}}
\def\qoc{I}
\def\eunscaled{{\cal E}}
\def\sep{{X}}
\def\surf{{\cal A}}
\def\sig{s}
\def\sigo{\Sigma}%
\def\sigoM{\sigma}
\def\sigoMhat{q}
\def\that{\omega}
\def\Sig{\Sigma}
\def\sigrel{\sigma}
\def\met{g}
\def\metinv{\bar{g}}
\def\metev{\gamma}
\def\prinrel{\zeta}
\def\prinsq{{\hat\prinrel}}
\def\prinsqmag{Z}
\def\defvol{{\cal W}}
\def\md{m}
\def\nd{b}
\def\ndp{b^{\prime}}
\def\nnd{\bar{b}}
\def\nndp{{\bar{b}}^{\prime}}
\def\shop{N}
\def\term{{\cal J}}
\def\coeff{{\cal B}}
\def\flat{F}
\def\height{H}
\def\bve{{\rm e}}
\def\contourinside{{\cal M}}
\def\contourclosed{{\partial\contourinside}}
\def\myalpha{{\Lambda}}
\def\vts{n}
\def\mom{P}


\section{%
\label{sec:intro}
Introduction and aims}
First conceived of and analyzed by Onsager~\cite{ref:Onsager-pub,ref:Onsager-coll} and
Feynman~\cite{ref:Feynman-pub,ref:Feynman-coll}, quantized vortices are topologically stable excitations of superfluid helium-four.  They are manifestations of quantum mechanics which owe their stability to the quantization law that the circulation of superflow obeys, leaving their imprint at arbitrarily large distances.  Vortex excitations play an essential role in determining the equilibrium properties of superfluid helium-four, especially in the mechanism via which superfluidity is lost as the temperature is raised through the \lq\lq lambda\rq\rq\ transition to the normal fluid state.  They also play an essential role in the mechanism via which superflow in toroid-shaped samples is dissipated.  For an in-depth account of the properties of quantized vortices, see Ref.~\cite{ref:Donnelly}.

When the space filled by the superfluid helium-four (i.e., the ambient space) is two- or three-dimensional, and when they are viewed at length-scales larger than their core size, quantized vortices are, respectively, the familiar, geometrical point-like vortices of two-dimensional superfluidity or line-like vortices of three-dimensional superfluidity.  The aim of the present paper is to examine some of the characteristic properties of quantized vortices that are familiar in two- and three-dimensional superfluid helium-four, and to determine how these topological excitations and their characteristic properties can be extended to situations in which the superfluid fills spaces having arbitrary numbers of dimensions.  For the case of three dimensions, these properties are analyzed in Ref.~\cite{ref:LLSP}.  Properties that we shall address here include the structure of quantized vortices and the flows around them, the motion that vortices inherit from such flows, and the flow energetics, all in a variety of settings.  In all dimensions, these excitations (again, when viewed at length-scales larger than their core size) continue to be geometrical structures having codimension two: their dimensionality grows with the dimension of the ambient space, always remaining two dimensions behind.  Nevertheless, we shall use the terminology \lq\lq quantized vortex\rq\rq\ (or, for simplicity, just \lq\lq vortex\rq\rq) for such excitations in all dimensions, not only two and three.

We have chosen to present our developments using the traditional language of vector and tensor calculus.  However, along the way we shall pause to mention how these developments appear when formulated in the language of exterior calculus and differential forms; see Ref.~\cite{ref:OnDiffForms}.  This language is in fact tailor-made for the questions at hand, and enables a highly economical development.  However, we have not relied on it, as there may be readers for whom it would require an unfamiliar level of abstraction.

This paper is organized as follows.
In Section~\ref{sec:setting}, we discuss the structure of vortices in superfluid helium-four in arbitrary dimensions of space, first from the perspectives of length-scales and topology, and then by developing the conditions that flows of a specific vorticial content must obey, in both the integral and the differential versions.
%
In Section~\ref{sec:EqFlowGiven}, we introduce considerations of kinetic energy, examining superflows of given vorticial content at equilibrium, and solving for the associated velocity fields of these flows.
In Section~\ref{sec:dynamics}, we analyze the dynamics that vortices inherit from the equilibrium flows associated with them.  Here, we specialize to asymptotically large, smooth vortices, for which the dynamics turns out to be a simple and clear reflection of the intrinsic and extrinsic geometry of the shapes of the vortices.  To illustrate our results, we consider the velocities of spherical and hyper-spherical vortices, as well as the small-amplitude excitations of planar and hyper-planar vortices.
In Section~\ref{sec:energcalc}, we return to issues of energetics, and determine the equilibrium kinetic energy of superflows containing individual vortices in terms of the geometry of the shapes of the vortices.  We illustrate this result for large, smooth---but otherwise arbitrarily shaped---vortices, for which a simple, asymptotic formula holds, and we give an elementary argument for this asymptotic formula.  Here, we also address the energetics of maximally symmetrically shaped spherical and hyper-spherical vortices, and we use our knowledge of the energies and velocities of the spherical and hyper-spherical vortices to determine their momenta and, hence, their energy-momentum relations.
In Section~\ref{ScalingDim}, we use scaling and dimensional analysis to obtain the aforementioned results, up to numerical factors and logarithmic dependencies on the short-distance cut-off, i.e., the core size of the vortices.  We end, in Section~\ref{sec:conclusions}, with some concluding remarks.
We have streamlined the presentation by relegating to a sequence of appendices much of the technical material.

\section{%
\label{sec:setting}
Structure of superfluid vortices in arbitrary dimensions}
\subsection{%
\label{sec:fixed-vortices}
General morphology of vortices }
We consider a large, $D$-dimensional, Euclidian, hyper-cubic volume $\vol$ of superfluid, with each point in $\vol$ being specified by its $D$-component position-vector $\pos$, the Cartesian components of which we denote by $\{\pos_{d}\}_{d=1}^{D}$.  On the faces of $\vol$ we suppose that all fields obey periodic boundary conditions.  We shall be concerned with flows of superfluid in $\vol$, and these we describe via their $D$-vector flow fields $\velg(\pos)$.  The phenomena that we shall be addressing occur on length-scales larger than the characteristic superfluid coherence- or healing-length $\vcs$, which determines the length-scale over which the superfluid density falls to zero as the centre of a vortex is approached (i.e., the vortex core size).

Provided we do indeed examine flows on length-scales larger than $\vcs$, and hence restrict our attention to long-wavelength, low-energy excitations of the superfluid, the {\it amplitude\/} $\vert\sfop(\pos)\vert$ of the complex scalar superfluid order parameter field $\sfop(\pos)$ can be taken to be a nonzero constant $\vert\sfop_{0}\vert$ away from the vortices.  In addition, these vortices can be idealized as being supported on vanishingly thin sub-manifolds of the ambient $D$-dimensional space.  The remaining freedom lies in the {\it phase factor\/} $\sfop(\pos)/\vert{\sfop(\pos)}\vert$ of $\sfop(\pos)$, and---except on vortex sub-manifolds, where it is not defined---this takes values on the unit circle in the complex plane.  Owing to the topology of the coset space (in this case, the unit circle) in which this phase factor resides, and specifically the fact that its fundamental homotopy group $\Pi_{1}$ is the additive group $\mathbb{Z}$ of the integers $\vts=0,\pm 1,\ldots$, these vortices are indeed topologically stable defects in the superfluid order~\cite{ref:mermin-RMP}.  The {\it strength\/} $\vts$ of any vortex is the number of times the phase factor winds through $2\pi$ as the vortex is encircled a single time.  For $D$-dimensional superfluids, and at the length-scales of interest, these vortices are supported on manifolds of dimension $D-2$, a combination that will occur frequently, so we shall denote it by $\Dsl$.  Said another way, the vortices reside on co-dimension-2 sub-manifolds of $D$-dimensional space, which means that we would have to add two further dimensions to \lq\lq fatten\rq\rq\ each vortex into a space-filling object.  In any dimension $D\ge 2$, there are one-dimensional closed paths that loop around the vortices.  As we shall discuss shortly, the circulation of the flow around such loops is quantized and nonzero, whereas the circulation around closed paths that do not loop around a vortex is zero.

\begin{figure}[htbp]
    \centering
	\includegraphics[width=.44\textwidth]{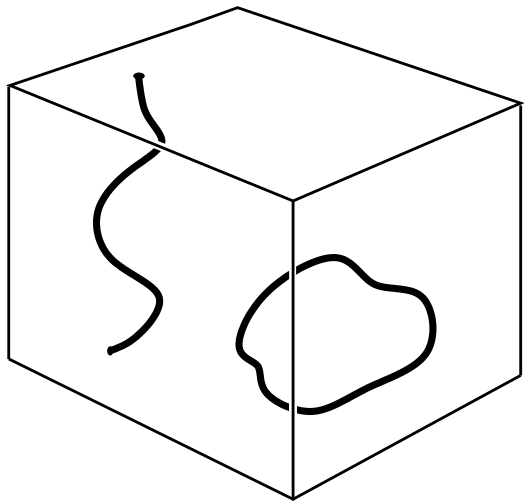}
	\caption{Vortices in three-dimensional superfluids.
    They form closed one-dimensional loops, possibly via the periodic
    boundary conditions.
}
\label{FIG:cube-with}
\end{figure}
In the familiar case of superfluids in three dimensions, the vortices are geometrical objects that are supported on one-dimen\-sional sub-manifolds (see Fig.~\ref{FIG:cube-with}).  The loops that encircle the vortices, along which the circulation of the flow is quantized, are also one-dimensional
objects~\footnote{The vortices themselves are closed loops that may or may not wind through
the boundaries of the sample, a possibility allowed by the  $D$-torus topology of the
sample.  This topology arises from the adopted periodic boundary conditions but will not
play a significant role here.}.
In the equally familiar case of superfluids in $D=2$, each vortex is a pair of opposing-circulation zero-dimensional objects (i.e., a structure that it usually called a vortex/anti-vortex pair).
The loops around one or other member of the pair, however, along which the circulation of the flow is quantized, remain one dimensional.
(Strictly speaking, in $D=2$ each vortex is supported on a pair of points, which is not a manifold, but we shall admit this abuse of nomenclature.)

Our main objectives are to explore the spatial structure and energetics of flows possessing a set of vortices at prescribed locations, as well as the motion that such flows confer on the vortices themselves.  The flows that we shall consider are states of thermodynamic equilibrium, but are maintained away from the no-flow state, $\velg=0$, because they are constrained to have a prescribed vorticial content.

\subsection{\label{sec:ampere}
Fixing the vorticial content of superflows}

The {\it circulation\/} $\vsk$ of the flow along any one-dimensional closed contour is defined to be the line-integral of the corresponding velocity field along that contour:
\begin{equation}
\vsk:=\oint\velg_{d}(\pos)\,d\pos_{d}\,.
\label{eq:circul-def}
\end{equation}
Here and throughout this paper, a summation from $1$ to $D$ is implied over repeated indices such as $d$, $d_{1}$, etc. The vorticial content of the flow can then be prescribed via the following condition.  For a vortex of strength $\vts$, the circulation $\vsk$ is quantized according to the number of times ${\cal N}$ the contour encircles the $\Dsl$ [i.e., $(D-2)$]-dimen\-sional sub-manifold supporting the vortex:
\begin{equation}
\kappa=(2\pi\hbar/M){\cal N}n,
\label{eq:ampere-int}
\end{equation}
where $\hbar$ is Planck's constant and $M$ is the mass of one of the particles whose condensation causes the superfluidity.  We adopt units in which
$\hbar/M$ has the value unity.  In the language of differential forms, the flow field is described by the 1-form $\velg(x)$, and the quantization of circulation~(\ref{eq:ampere-int}) is expressed as
\begin{equation}
\oint\velg=2\pi{\cal N}n.
\label{eq:ampere-int-DF}
\end{equation}
We shall be concerned with vortices of unit strength (i.e., $\vts=1$).  Owing to the linearity of the theory, results for higher-strength vortices can be straightforwardly obtained from those for unit-strength vortices via elementary scalings with $\vts$ (i.e., velocities scale linearly with $\vts$; energies scale quadratically).

For a flow $\velg$ to have the prescribed vorticial content, it must be expressible as a sum of two contributions:
\begin{enumerate}
\item
a real vector field $\vel(\pos)$ that is singular on any vortex-supporting sub-manifolds, and has the appropriate quantized circulation around them, and thus its contribution to the flow includes the flow's vorticial part; and
\item
the gradient of a real, scalar phase field $\Phase(\pos)$ that is smooth and {\it single-valued\/} throughout $\vol$.  Owing to the latter property, this component of the flow is capable of contributing only to the irrotational (i.e., circulation-free) aspects of the flow.
\end{enumerate}
Thus, we consider flows of the form
\begin{equation}
\velg_{d}(\pos)=\vel_{d}(\pos)+\nabla_{d}\,\Phase(x),
\label{eq:flow-break}
\end{equation}
in which the gradient $\nabla_{d}$ denotes the partial derivative $\partial/\partial \pos_{d}$ with respect to the $d^{\rm th}$ spatial coordinate $\pos_{d}$.
(The $\nabla_{d}\,\Phase$ term carries a dimensional factor of $\hbar/M$, which we set to unity, earlier.)\thinspace\
In the language of differential forms, Eq.~(\ref{eq:flow-break}) is expressed as
\begin{equation}
\velg=\vel+d\Phase,
\label{eq:flow-break-DF}
\end{equation}
where $d\Phase$ is the exact 1-form that results from taking the exterior derivative of the 0-form (i.e., the function) $\Phase$.

To ensure that the vorticial part $\vel$ of the flow $\velg$ has the appropriate circulation on loops surrounding any vortex-supporting sub-manifolds, it must obey the following inhomogeneous partial differential equation:
\begin{equation}
\epsilon_{d_{1}\cdots d_{D}}
\nabla_{d_{D-1}}\vel_{d_{D}}(\pos)
=2\pi\source_{d_{1}\cdots d_{\Dsl}}(\pos),
\label{eq:ampere}
\end{equation}
where $\epsilon_{d_{1}\cdots d_{D}}$ is the completely skew-symmetric $D$-dimensional Levi-Civita symbol.  Equation~(\ref{eq:ampere}) is a natural generalization to $D$-dimensions of the static version of the three-dimensional differential expression of the Amp\`ere-Maxwell law relating the magnetic field (the analogue of $\vel$) to the electric current density (the analogue of $\source$), which in suitable units reads
\begin{equation}
{\rm curl}\,{\bf V}
=2\pi{\bf\source}.
\label{eq:ampere-3}
\end{equation}
An account of the analogy between superfluidity and magnetostatics for the case of three dimensions is given in Ref.~\cite{ref:Kleinert}.

The source field in Eq.~(\ref{eq:ampere}), $\source$, is a completely skew-symmetric, rank $\Dsl$ tensor field, which obeys the divergencelessness condition $\partial_{d_{1}}\source_{d_{1}\cdots d_{\Dsl}}(\pos)=0$, i.e., the consistency condition that follows, e.g., from the application of $\partial_{d_{1}}$ to Eq.~(\ref{eq:ampere}).
(We discuss this point further in \ref{ap:derivation-stokes-D}.)\thinspace\
This field $\source$ encodes the singularity in the density of the vorticity associated with any $\Dsl$-dimensional vortices in the superflow.
The factor of $2\pi$ in Eq.~(\ref{eq:ampere}) is extracted to ensure that unit-strength sources gives rise to unit quanta of circulation.
That $\vel$ must obey Eq.~(\ref{eq:ampere}) follows from the integral form of this law, Eq.~(\ref{eq:ampere-int}), which can be obtained from Eq.~(\ref{eq:ampere}) via the $D$-dimensional version of Stokes' theorem.
The right-hand side of Eq.~(\ref{eq:ampere-int}) is the flux of the source $\source(\pos)$
through the corresponding closed contour.
(In its integral form, the three-dimensional Amp\`ere-Maxwell law dictates that the
line-integral of the magnetic field along a loop in three-dimensional space has a value
determined by the flux of the electric current threading the loop.)\thinspace\
The skew-symmetric tensor nature of the source field $\source$, present for $D\ge 4$, is a natural generalization of the vectorial nature of $\source$ present in $D=3$, and reflects the corresponding higher-than-one-dimensional geometry of the sub-manifold that supports the vortex, as we shall discuss shortly.

In the language of differential forms, Eq.~(\ref{eq:ampere}) is expressed as an equality of $\Dsl$-forms, i.e.,
\begin{equation}
\hodge\,d\vel=2\pi\source,
\label{eq:ampere-DF}
\end{equation}
where $d\vel$ is the 2-form that results from taking the exterior derivative of $\vel$, the form $\hodge\,d\vel$ is the $\Dsl$-form dual to it that results from applying the (Eucidian) Hodge-$\hodge$ operator to it, and $\source$ is the source $\Dsl$-form.  The dual of this equation, which is an equality of 2-forms obtained by applying $\hodge$ to it, reads
\begin{equation}
d\vel=\hodge\,(2\pi\source).
\label{eq:ampere-DF-dual}
\end{equation}
In the language of differential forms, the divergencelessness of the source $\source$, mentioned shortly after Eq.~(\ref{eq:ampere-3}), amounts to $\source$ being a closed form, i.e., $d\source=0$.

\begin{figure}[htbp]
    \centering
	\includegraphics[width=.38\textwidth]{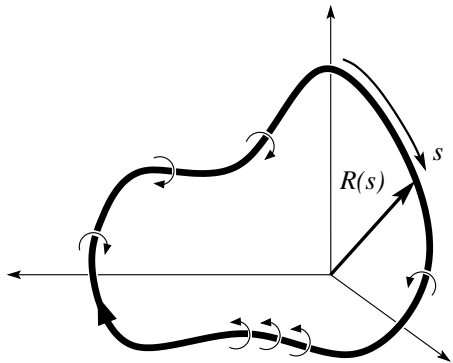}
	\caption{A loop vortex in a three-dimensional superfluid specified by $\dfc(\sig)$
    with arc-length parameter $\sig$, showing the direction of the source
    \lq\lq electrical current\rq\rq\ along the loop and the sense of the superflow
    encircling the loop.}
\label{FIG:1-loop-vortex}
\end{figure}

\begin{figure}[htbp]
    \centering
    \includegraphics[width=.40\textwidth]{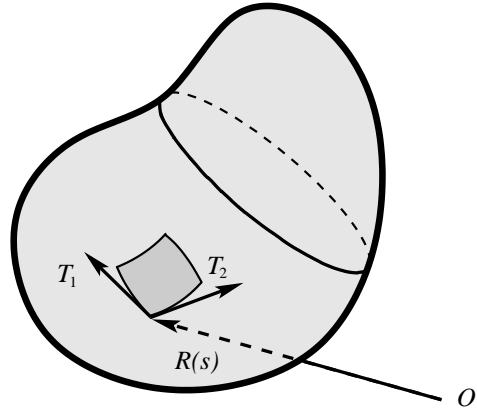}
	\caption{A two-dimensional vortex sub-manifold in a four-dimensional superfluid.
    $T_1$ [$=\newpart_1 \dfc(\sig)$] and $T_2$ [$=\newpart_2 \dfc(\sig)$]
    are two coordinate tangent vectors to the sub-manifold at the point $\dfc(\sig)$.
    These tangent vectors define a \lq\lq patch\rq\rq\ of the sub-manifold.}
\label{FIG:tan-vectors}
\end{figure}

What value should the source field $\source$ have if the $D$-dimensional flow is to contain a single, unit-strength, quantized vortex?  Recall that for $D=3$ (so that $\Dsl=1$), the vortex is one-dimensional and supported on the curve ${\bf\pos}={\bf\dfc}(\sig_{1})$, parameterized by the single variable $\sig_{1}$ (see Fig.~\ref{FIG:1-loop-vortex}).  In this case, the source takes the value
\begin{equation}
{\bf\source}({\bf\pos})=
\int d\sig_{1}\,
\frac{d{\bf R}}{d\sig_{1}}\,
\,\dirac^{(3)}\!
\left({\bf\pos}-{\bf\dfc}(\sig_{1})\right),
\label{eq:current-3D}
\end{equation}
associated with the analog of a unit-strength electrical current flowing along the curve and supported on it.  The sign of this analogue of the current (i.e., its direction around the loop) determines the sense in which the superflow swirls around the vortex core, via a right-hand rule.
Observe that the geometry of the line enters not only in terms of the curve ${\bf R}(\sig_{1})$
on which the vortex is supported, but also via the tangent vector to this curve,
${d{\bf R}}/{d\sig_{1}}$.
The geometrically natural extension of the source to arbitrary dimension $D$ involves:
\begin{enumerate}
\item
concentrating the source on the appropriate $\Dsl$-dimensional sub-manifold;
\item
specifying the shape of this sub-manifold, via $\pos=\dfc(\sig)$, where $\sig$ denotes the set of the $\Dsl$ variables $\{\sig_{1},\sig_{2},\ldots,\sig_{\Dsl}\}$ that, when they explore their domains, trace out the idealized core of the vortex; and
\item
endowing the source with appropriate geometrical content, via the natural generalization of the tangent vector.
\end{enumerate}
This generalization is provided by the oriented, $\Dsl$-dimensional (tensorial) volume
$\varepsilon_{\inin_{1}\cdots\inin_{\Dsl}}
(\newpart_{\inin_{1}}\dfc_{d_{1}})
\cdots
(\newpart_{\inin_{\Dsl}}\dfc_{d_{\Dsl}})$
(where $\newpart_{\inin}$ denotes the partial derivative with respect to the $\inin^{\rm th}$ independent variable that parameterizes the vortex sub-manifold, in this case, $\partial/\partial\sig_{\inin}$) of the parallelepiped constructed from
$\{\partial\dfc/\partial\sig_{\inin}\}_{\inin=1}^{\Dsl}$,
i.e., the set of $\Dsl$ linearly independent, coordinate tangent-vectors to the sub-manifold at $\dfc(\sig)$.
Together with the measure $d^{\Dsl}\!\sig$, this volume plays the role of an infinitesimal segment of vortex line (in $D=3$) or patch of vortex area (in $D=4$, as sketched in Fig.~\ref{FIG:tan-vectors}).
Here, $\varepsilon_{\inin_{1}\cdots\inin_{\Dsl}}$ is the completely skew-symmetric $\Dsl$-dimensional Levi-Civita symbol, in contrast with the $D$-dimensional $\epsilon_{d_{1}\cdots d_{D}}$, which we defined shortly after Eq.~(\ref{eq:ampere}).

\begin{figure}[htbp]
    \centering
	\includegraphics[width=.48\textwidth]{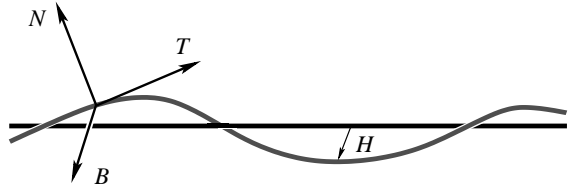}
	\caption{A segment of a vortex in a three-dimensional superfluid,
    showing the tangent vector $d\dfc/d\sig$ and two normal vectors
    $\shop^1$ and $\shop^2$.}
\label{FIG:1d-vortex}
\end{figure}
Thus, for the case of a $\Dsl$-dimensional vortex in a $D$-dimensional flow, the source takes the form
\begin{equation}
\source_{d_{1}\cdots d_{\Dsl}}(\pos)=
\int d^{\Dsl}\sig\,
\varepsilon_{\inin_{1}\cdots\inin_{\Dsl}}
(\newpart_{\inin_{1}}\dfc_{d_{1}})
\cdots
(\newpart_{\inin_{\Dsl}}\dfc_{d_{\Dsl}})\,
\dirac^{(D)}\!\left(\pos-\dfc(\sig)\right)\!,
\label{eq:current-allD}
\end{equation}
where summations from $1$ to $\Dsl$ are implied over repeated indices $\inin_{1},\inin_{2}$, etc.
We show in \ref{ap:derivation-stokes-D} that this form of the source does indeed describe a
generalized vortex of unit vorticity, in the sense that the circulation of the superflow computed
on any one-dimensional loop that surrounds the source a single time is unity (up to a sign that
depends on the sense in which the loop is traversed).
Furthermore, this form of the source obeys the condition of being closed, i.e.,
$\newpart_{d_{1}}\source_{d_{1}\cdots d_{D-2}}(\pos)=0$, as we also show
in \ref{ap:derivation-stokes-D}.
In the language of differential forms, Eq.~(\ref{eq:current-allD}) is expressed as
\begin{equation}
\source(\pos)=
\int d^{\Dsl}\!\sig\,\,
\dirac^{(D)}\!\left(\pos-\dfc(\sig)\right)\,
(\newpart_{1}\dfc)\wedge
(\newpart_{2}\dfc)\wedge
\cdots
\wedge(\newpart_{\Dsl}\dfc).
\label{eq:current-allD-forms}
\end{equation}

\begin{figure}[htbp]
    \centering
	\includegraphics[width=.48\textwidth]{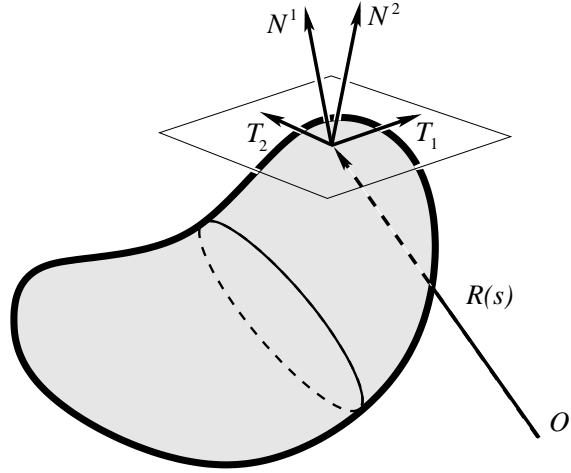}
	\caption{A two-dimensional vortex sub-manifold in a four-dimensional superfluid.
    Shown are two tangent vectors, $T_1$ and $T_2$, and two normal vectors,
    $\shop^1$ and $\shop^2$, at the point $\dfc(\sig)$.}
\label{FIG:2d-vortex}
\end{figure}
It is also useful to consider the dual of the source, viz.,
the completely skew-symmetric rank-2 field
\begin{eqnarray}
&&\sourcedual_{d_{D-1}d_{D}}(\pos):=
\frac{1}{\Dsl !}\,
\epsilon_{d_{1}d_{2}{\cdots}d_{\Dsl}d_{D-1}d_{D}}\,
\source_{d_{1}\cdots d_{\Dsl}}(\pos)
\nonumber\\
&&\quad=
\frac{\epsilon_{d_{1}{\cdots}d_{D-1}d_{D}}}{\Dsl !}
\int d^{\Dsl}\!\sig\,
\varepsilon_{\inin_{1}\cdots\inin_{\Dsl}}
(\newpart_{\inin_{1}}\dfc_{d_{1}})
\cdots
(\newpart_{\inin_{\Dsl}}\dfc_{d_{\Dsl}})\,
\dirac^{(D)}\left(\pos-\dfc(\sig)\right)\!.
\label{eq:current-allD-dual}
\end{eqnarray}
Let us pause to analyze the factor
\begin{equation}
[\Dsl !]^{-1}\,
\epsilon_{d_{1}{\cdots}d_{D-1}d_{D}}\,
\varepsilon_{\inin_{1}\cdots\inin_{\Dsl}}\,
(\newpart_{\inin_{1}}\dfc_{d_{1}})
\cdots
(\newpart_{\inin_{\Dsl}}\dfc_{d_{\Dsl}})
\label{eq:geom-factor}
\end{equation}
from the integrand of Eq.~(\ref{eq:current-allD-dual}), and hence to introduce some useful geometrical ideas.
Being a skew-symmetrized (over the {\it spatial\/} indices $d_{1}$, $d_{2}$ etc.)~product of the set of vectors $\{\newpart_{\inin}\dfc\}_{\inin=1}^{\Dsl}$
tangent to the sub-manifold at the point $\dfc(\sig)$, the skew-symmetrization over indices associated with the sub-manifold coordinates (via $\varepsilon_{\inin_{1}\cdots\inin_{\Dsl}}$) is elementary, yielding a factor of $\Dsl!$, so that the factor can equally well be written as
\begin{equation}
\epsilon_{d_{1}\cdots d_{D}}\,
(\newpart_{1}\dfc_{d_{1}})
 \cdots
(\newpart_{\Dsl}\dfc_{d_{\Dsl}}).
\label{FIG:firstNs}
\end{equation}
Being orthogonal to the plane tangent to the manifold (i.e., the plane spanned by the vectors
$\{\newpart_{\inin}\dfc\}_{\inin=1}^{\Dsl}$),
this factor can be expressed in terms of a skew-symmetric product of {\it any pair\/}, $\shop^{1}$ and $\shop^{2}$,
of unit vectors that are mutually orthogonal and orthogonal to the tangent plane
(see
Fig.~\ref{FIG:1d-vortex} for an example in $D=3$
and
Fig.~\ref{FIG:2d-vortex} for an example in $D=4$),
i.e., as
\begin{equation}
\big(
\shop_{d_{D-1}}^{1}\shop_{d_{D   }}^{2}-
\shop_{d_{D   }}^{1}\shop_{d_{D-1}}^{2}
\big)\,\coeff,
\label{FIG:SS-Ns}
\end{equation}
in terms of a coefficient $\coeff$.
To determine $\coeff$ (up to a sign associated with orientation), we take the formula
\begin{equation}
\big(
\shop_{d_{D-1}}^{1}\shop_{d_{D   }}^{2}-
\shop_{d_{D   }}^{1}\shop_{d_{D-1}}^{2}
\big) \coeff
=
\epsilon_{d_{1}\cdots d_{D}}
(\newpart_{1}\dfc_{d_{1}})
\cdots
(\newpart_{\Dsl}\dfc_{d_{\Dsl}}),
\end{equation}
and fully contract each side with itself to obtain
\begin{eqnarray}
\coeff^{2}&=&{\scriptstyle{\frac{1}{2}}}
\epsilon_{d_{1}\cdots d_{\Dsl}d_{D-1}d_{D}}\,
(\newpart_{1}\dfc_{d_{1}})
\cdots
(\newpart_{\Dsl}\dfc_{d_{\Dsl}})\,
\epsilon_{\bd_{1}\cdots \bd_{\Dsl}d_{D-1}d_{D}}\,
(\newpart_{1}\dfc_{\bd_{1}})
\!\cdots\!
(\newpart_{\Dsl}\dfc_{\bd_{\Dsl}})
\nonumber
\\
\noalign{\smallskip}
&=&(\delta_{d_1 \bd_1}\delta_{d_2 \bd_2}\cdots\delta_{d_\Dsl \bd_\Dsl}-
\delta_{d_1 \bd_2}\delta_{d_2 \bd_1}\cdots\delta_{d_\Dsl \bd_\Dsl}+\cdots)
\nonumber\\
&&\qquad\qquad\qquad\times
(\newpart_{1}\dfc_{d_{1}})
\!\cdots\!
(\newpart_{\Dsl}\dfc_{d_{\Dsl}})
(\newpart_{1}\dfc_{\bd_{1}})
\!\cdots\!
(\newpart_{\Dsl}\dfc_{\bd_{\Dsl}})
\nonumber
\\
\noalign{\smallskip}
&=&
+(\newpart_{1}\dfc_{d_{1}})(\newpart_{1}\dfc_{d_{1}})\,
(\newpart_{2}\dfc_{d_{2}})(\newpart_{2}\dfc_{d_{2}})
\cdots
(\newpart_{\Dsl}\dfc_{d_{\Dsl}})(\newpart_{\Dsl}\dfc_{d_{\Dsl}})
\nonumber\\
&&
-(\newpart_{1}\dfc_{d_{1}})(\newpart_{2}\dfc_{d_{1}})\,
(\newpart_{2}\dfc_{d_{2}})(\newpart_{1}\dfc_{d_{2}})
\cdots
(\newpart_{\Dsl}\dfc_{d_{\Dsl}})(\newpart_{\Dsl}\dfc_{d_{\Dsl}})+
\cdots
\nonumber
\\
&=&{\rm det}\,\met,
\end{eqnarray}
where $\met_{\inin^{\prime}\inin}$ are the components of the {\it{induced metric}} $\met$ on the vortex sub-manifold,
also known as the {\it first fundamental form\/}~\cite{ref:fundamental-forms}, and are defined via
\begin{equation}
\met_{\inin^{\prime}\inin}(\sig):=
\frac{\partial\dfc_{d}(\sig)}{\partial\sig_{\inin^{\prime}}}
\frac{\partial\dfc_{d}(\sig)}{\partial\sig_{\inin}}.
\label{eq:def-ind-metric}
\end{equation}
Thus, for the geometric factor~(\ref{eq:geom-factor}) we obtain
\begin{equation}
\sqrt{\det g}\,
\big(\shop_{d_{D-1}}^{1}\shop_{d_{D}}^{2}-
\shop_{d_{D-1}}^{2}\shop_{d_{D}}^{1}\big),
\label{eq:geom-factor2}
\end{equation}
and for the components of the dual source field $\hodge\,\source$ we obtain
\begin{eqnarray}
\sourcedual_{d_{D-1}d_{D}}(\pos)&=&
\int d^{\Dsl}\!\sig\,\sqrt{\det g}
\nonumber \\
&&\quad \times
\big(
\shop_{d_{D-1}}^{1}\shop_{d_{D  }}^{2}-
\shop_{d_{D  }}^{1}\shop_{d_{D-1}}^{2}
\big) \,
\dirac^{(D)}\!\left(\pos-\dfc(\sig)\right).
\end{eqnarray}
The measure $d^{\Dsl}\sig\,\sqrt{\det g}$ is the infinitesimal area element on the vortex sub-manifold $\dfc(\sig)$.
In the language of differential forms, Eq.~(\ref{eq:current-allD-dual}) is expressed as
\begin{equation}
\sourcedual(\pos)=\int d^{\Dsl}\!\sig\,\sqrt{\det g}\,
\big(\shop^{1}\wedge\shop^{2}\big)\,
\dirac^{(D)}\!\left(\pos-\dfc(\sig)\right).
\end{equation}

\section{%
Equilibrium superflows of a given vorticial content
\label{sec:EqFlowGiven}}
In this section, we introduce energetic considerations and show that---provided we adopt a suitable choice of gauge for the {\it{rotational}} part $\vel$ of the flow---the {\it{irrotational}} contribution to the flow $\nabla\Phase$ is zero in the equilibrium state.  After that, introduce a convenient gauge potential $\gp$, and use it to solve for the flow in the presence of a given vorticial content, the latter being specified by a suitable source term.  The gauge potential is a rank-$\Dsl$, skew-symmetric tensor, precisely analogous to the vector potential of three-dimensional magnetostatics.

\subsection{\label{sec:flow-energetics}
Minimization of superflow kinetic energy}
We take the energy of the system to be entirely associated with the kinetic energy of superflow.  As we are addressing flows possessing a {\it prescribed\/} set of vortices, the energy associated with the absence of superfluidity in the cores of the vortices will not play a significant role.  We assume that the mass-density of the superfluid $\mdc$ is constant, i.e., we take the fluid to be incompressible.  The energy $\mdc E$ associated with the flow $\velg$ can therefore be specified in terms of the functional
\begin{equation}
E:=\frac{1}{2}\int_{\vol}d^{D}\pos\,
\velg_{d}(\pos)\,\velg_{d}(\pos),
\label{eq:KEintegral}
\end{equation}
which, for convenience, we shall refer to as the energy, despite the missing factor of $\mdc$.

To determine the equilibrium flow in the presence of some prescribed vorticial content, we regard $\vel$ in Eq.~(\ref{eq:flow-break}) as fixed and minimize $E$ with respect to $\Phi$. The corresponding stationarity condition reads:
\begin{eqnarray}
0&=&
\frac{\delta E}{\delta\Phase(\pos)}
=\frac{\delta}{\delta\Phase(\pos)}
\frac{1}{2}
\int_{\vol}d^{D}\bpos\,
[\vel_{d}(\bpos)+\nabla_{d}\Phase(\bpos)]\,
[\vel_{d}(\bpos)+\nabla_{d}\Phase(\bpos)]
\\
\noalign{\medskip}
&=&
  -\nabla_{d}\big(\nabla_{d}\,\Phase(\pos)-\vel_{d}(\pos)\big),
\label{eq:minimizer}
\end{eqnarray}
where we have used the periodic boundary conditions on $\vel$ and $\Phase$ to eliminate boundary terms, and thus arrive at an overall condition of divergencelessness on the flow $\velg$.  Now, $\vel$ obeys Eq.~(\ref{eq:ampere}), and is therefore not unique, being adjustable by any gradient of a scalar.  This gauge freedom permits us to demand that the rotational part of the flow itself be divergenceless, i.e.,
\begin{equation}
\nabla_{d}\vel_{d}=0,
\label{eq:divergence}
\end{equation}
and we shall impose this demand.  Thus, from Eq.~(\ref{eq:minimizer}) we see that the minimizer $\Phase$ is harmonic, i.e. $\nabla_{d}\nabla_{d}\,\Phase=0$, and subject to periodic boundary conditions.  Thus, $\Phase$ is a constant, and does not contribute to the equilibrium flow $\velg$, which is given by $\vel$ alone, the latter obeying  Eqs.~(\ref{eq:ampere}) and (\ref{eq:divergence}).  Thus, the flow has energy $(\mdc/2)\int d^{D}\pos\,\vel_{d}\vel_{d}$.  (Minimization, and not just stationarity, of the energy follows directly from the positive semi-definiteness of the second variation of $E$.)

\subsection{\label{sec:fixeqm}
Fixing the complete flow: equilibrium state at given vorticial content}

To solve Eq.~(\ref{eq:ampere}) in terms of the source and, more particularly, the location of the singular surface $\dfc(\sig)$, we take advantage of the divergencelessness of $\vel$, Eq.~(\ref{eq:divergence}).  This allows us to introduce a gauge field $\gp$, which is a completely skew-symmetric rank-$\Dsl$ tensor field, in terms of which $\vel$ is given by
\begin{equation}
\vel_{d_{D}}(\pos)=
\epsilon_{d_{1}\cdots d_{D}}\,
\nabla_{d_{D-1}}
\gp_{d_{1}\cdots d_{\Dsl}}(\pos).
\label{eq:vel-curl}
\end{equation}
We may subject this gauge field to a gauge condition, and the one that, for convenience, we adopt, is the transverse condition:
\begin{equation}
\nabla_{d_{\Dsl}}\,
\gp_{d_{1}\cdots d_{\Dsl}}(\pos)=0.
\label{eq:lorentz}
\end{equation}
In the language of differential forms Eqs.~(\ref{eq:vel-curl}) and (\ref{eq:lorentz}) are expressed as
\begin{equation}
V=\hodge\,\,dA
\quad{\rm and}\quad
dA=0.
\end{equation}

To solve for $\gp$, we insert $\vel$, expressed in terms of $\gp$, into Eq.~(\ref{eq:ampere}), to obtain
\begin{eqnarray}
\epsilon_{d_{1}\cdots d_{D}}
\nabla_{d_{D-1}}
\epsilon_{\bd_{1}\cdots\bd_{D-1}d_{D}}
\nabla_{\bd_{D-1}}
\gp_{\bd_{1}\cdots\bd_{\Dsl}}
&=&
2\pi\source_{d_{1}\cdots d_{\Dsl}}.
\end{eqnarray}
We then contract the two $\epsilon$ symbols, which yields a [$(D-1)!$ termed] sum of products of $(D-1)$ Kronecker deltas, signed with the signature of the permutation $\perm$, i.e.,
\begin{equation}
\epsilon_{d_{1}\cdots d_{D}}\,
\epsilon_{\bd_{1}\cdots\bd_{D-1}d_{D}}=
\sum\nolimits_{\perm}
{{\rm sgn}(\perm})\,
\delta_{d_{1}\perm(\bd_{1})}\cdots\delta_{d_{D-1}\perm(\bd_{D-1})},
\end{equation}
and employ the skew symmetry of $\gp$ and the gauge condition~(\ref{eq:lorentz}) to arrive at the generalized Poisson equation, viz.,
\begin{equation}
\Dsl!\,\,\nabla_{d}\nabla_{d}\,\gp_{d_{1}\cdots d_{D-2}}(\pos)=
2\pi\source_{d_{1}\cdots d_{\Dsl}}(\pos).
\label{eq:Poisson}
\end{equation}
By virtue of the adopted gauge condition, Eq.~(\ref{eq:Poisson}) has the cartesian components of $\gp$ uncoupled from one another.  Lastly, we introduce the $D$-dimensional Fourier transform and its inverse, defined as
\begin{eqnarray}
f(q   )&=&\int_{\vol} d^{D}\pos \,{\rm e}^{-iq\cdot\pos}\,f(\pos),\\
f(\pos)&=&\int \dbar^{D}q       \,{\rm e}^{ iq\cdot\pos}\,f(q)   ,
\end{eqnarray}
where $q$ is a $D$-vector and
$\dbar^{D}q:=d^{D}q/(2\pi)^D$,
and apply it to Eq.~(\ref{eq:Poisson}) to take advantage of the translational invariance of the Laplace operator.  (We are implicitly taking the large-volume limit, and thus the inverse Fourier transform involves an integral over wave-vectors rather than a summation.)\thinspace\  Thus, we arrive at an algebraic equation relating the Fourier transforms of the gauge field $\gp$ and the source $\source$, which we solve to obtain
\begin{equation}
\gp_{d_{1}\cdots d_{\Dsl}}(q)=
-\frac{2\pi}{\Dsl!}\frac{\source_{d_{1}\cdots d_{\Dsl}}(q)}{q\cdot q}.
\end{equation}
Inverting the Fourier transform, $\gp(q)$, yields the real-space gauge field
\begin{equation}
\gp_{d_{1}\cdots d_{\Dsl}}(x)=
-\frac{2\pi}{\Dsl !}\int\dbar^{D}\!q\,
\frac{{\rm e}^{iq\cdot\pos}}{q\cdot q}
\source_{d_{1}\cdots d_{\Dsl}}(q),
\end{equation}
from which, via Eq.~(\ref{eq:vel-curl}), we obtain the Fourier representation of the real-space velocity field,
\begin{equation}
\label{eq:velfieldsource}
\vel_{d_{D}}(x)=
-\frac{2\pi\,}{\Dsl !}\,
\epsilon_{d_{1}\cdots d_{D}}\,
\int\dbar^{D}\!q\,
\frac{iq_{d_{D-1}}\,{\rm e}^{iq\cdot\pos}}{q\cdot q}\,
\source_{d_{1}\cdots d_{\Dsl}}(q).
\end{equation}
For a source $\source$ given by a single vortex, as in Eq.~(\ref{eq:current-allD}), we have the following, more explicit form for the velocity field:
\begin{eqnarray}
\vel_{d_{D}}(x)&=&
-\frac{2\pi}{\Dsl\,!}\,
\epsilon_{d_{1}\cdots d_{D}}
\int\dbar^{D}\!q\,
\frac{iq_{d_{D-1}}}{q\cdot q}
{\rm e}^{-iq\cdot\pos}
\nonumber
\\
&&\qquad\qquad\times
\int d^{\Dsl}\sig\,
\varepsilon_{\inin_{1}\cdots\inin_{D-2}}
(\newpart_{\inin_{1}}\dfc_{d_{1}})
\cdots
(\newpart_{\inin_{\Dsl}}\dfc_{d_{\Dsl}})\,
{\rm e}^{iq\cdot\dfc(\sig)}.
\label{eq:vel-formula}
\end{eqnarray}
As shown in \ref{ap:velocity-int}, the $q$ integration may be performed, yielding the result
\begin{eqnarray}
\vel_{d_{D}}(x)&=&
\frac{1}{2\,\pi^{\Dsl/2}}\,\frac{\Gamma(\Dsl/2)}{(\Dsl-1)!}\,
 \epsilon_{d_{1}\cdots d_{D}}
\nonumber
\\
&&\qquad\times
\int d^{\Dsl}\sig\,
\varepsilon_{\inin_{1}\cdots\inin_{\Dsl}}
(\newpart_{\inin_{1}}\dfc_{d_{1}})
\cdots
(\newpart_{\inin_{\Dsl}}\dfc_{d_{\Dsl}})\,
\frac{\big(\pos-\dfc(\sig)\big)_{d_{D-1}}}{\!\!\!\!\!\!\big\vert{\pos-\dfc(\sig)}\big\vert^{D}},
\label{eq:vel-formula-surf}
\end{eqnarray}
in which $\Gamma(z)$ is the standard Gamma function~\cite{ref:Abram-handbook}.

Formula~(\ref{eq:vel-formula-surf}) for the velocity field $\vel$ gives the {\it structure\/}
of the superflow at any point $\pos$ in the superfluid in terms of the shape $\dfc$ of a vortex.
In the following sections, we shall deploy this formula to address the following issues:
the velocity conferred by this flow on vortices themselves,
the velocity of hyper-spherical vortices,
the dispersion relation for small deformations of hyper-planar vortices,
the superflow kinetic energy for vortices of arbitrary shape,
and the energy-momentum relation for hyper-spherical vortices.

\section{%
\label{sec:dynamics}
Dynamics of vortices in arbitrary dimensions}

\subsection{\label{sec:velocity}
Motion of vortices: asymptotics and geometry}
We have already established the flow field in the presence of a vortex, Eq.~(\ref{eq:vel-formula}).  We now examine the velocity $\velv(\sigo)$ that this flow confers on points $\pos=\dfc(\sigo)$ {\it on\/} the vortex itself:
\begin{eqnarray}
&&\velv_{d_{D}}(\sigo)
:=
\vel_{d_{D}}(x)\Big\vert_{x=\dfc(\sigo)}=
\frac{1\,}{2\pi^{\Dsl/2}}
\frac{\Gamma\left(\Dsl/2\right)}{(\Dsl-1)!}\,
\epsilon_{d_{1}\cdots d_{D}}
\nonumber
\\
&&\qquad\quad\times
\int d^{\Dsl}\sig\,
\varepsilon_{\inin_{1}\cdots\inin_{\Dsl}}\,
\newpart_{\inin_{1}}\dfc_{d_{1}}(\sig)
\cdots
\newpart_{\inin_{\Dsl}}\dfc_{d_{\Dsl}}(\sig)\,
\frac{\big(\dfc(\sigo)-\dfc(\sig)\big)_{d_{D-1}}}
{\big\vert{\dfc(\sigo)-\dfc(\sig)}\big\vert^{D}}.
\label{eq:vel-local}
\end{eqnarray}
Observe that the integral in Eq.~(\ref{eq:vel-local}) is logarithmically divergent; physically, it is regularized via the short-distance cut-off $\vcs$.

We now determine the leading contribution to the velocity in a regime in which the characteristic linear dimension of the vortex and the radii of curvature of the sub-manifold supporting the vortex are large, compared with $\vcs$.  In this limit, the dominant contribution to the integral is associated with small values of the denominator.  As we show in \ref{ap:vortex-vel}, the asymptotic behaviour of $\velv(\sigo)$ is given by
\begin{eqnarray}
&&
\velv_{d_{D}}(\sigo)\approx
\frac{1}{\Dsl !}\,\frac{\Gamma(\Dsl/2)}{2 \pi^{\Dsl/2}}
\frac{1}{\sqrt{{\rm det}\,\met}}\,
\surf_{\Dsl}
\ln(L/\vcs)\,
\epsilon_{d_{1}\cdots d_{D}}\,
\varepsilon_{\inin_{1}\cdots\inin_{\Dsl}}
\nonumber\\&&\qquad\qquad\quad\times
\Big\{
-\frac{1}{2}
 (\newpart_{\inin_{1}}\dfc_{d_{1}})\,
 \cdots
 (\newpart_{\inin_{\Dsl}}\dfc_{d_{\Dsl}})\,
 (\newpart_{\nd}\newpart_{\ndp}\dfc_{d_{D-1}})\,
 \metinv_{\nd\ndp}
 \nonumber\\&&\qquad\qquad\quad\phantom{\times\Big\{\,\,\,}
-(\newpart_{\ndp}\dfc_{d_{D-1}})\,
 (\newpart_{\inin_{1}}\dfc_{d_{1}})\,
 \cdots
 (\newpart_{\nd}\newpart_{\inin_{\md}}\dfc_{d_{\md}})\,
 \cdots
 (\newpart_{\inin_{\Dsl}}\dfc_{d_{\Dsl}})\,
 \metinv_{\nd\ndp}
 \nonumber\\&&\qquad\qquad\quad\phantom{\times\Big\{\,\,\,}
+\frac{1}{2}
 (\newpart_{\nd}\dfc_{\deflambda})\,
 (\newpart_{\nnd}\newpart_{\nndp}\dfc_{\deflambda})\,
 (\newpart_{\inin_{1}}\dfc_{d_{1}})\,
 \cdots
 (\newpart_{\inin_{\Dsl}}\dfc_{d_{\Dsl}})\,
 (\newpart_{\ndp}\dfc_{d_{D-1}})\,
 \nonumber\\&&\qquad\qquad\qquad\qquad\phantom{\times\Big\{\,\,\,}
 \times
 \big(\metinv_{\nd\nnd}\metinv_{\ndp\nndp}+
       \metinv_{\nd\nndp}\metinv_{\nnd\nndp}+
       \metinv_{\nd\ndp}\metinv_{\nnd\nndp}\big)
\Big\},
\label{eq:vel-local-approx}
\end{eqnarray}
in which all terms, such as gradients of $\dfc$, the metric tensor $\met$ and the inverse $\metinv$ of the metric tensor, are evaluated at the point $\Sig$ on the vortex sub-manifold, and the repeated indices $\nd$ etc.\ are summed from $1$ to $\Dsl$.  The factor $\surf_{\Dsl}$ is the surface area of a $\Dsl$-dimensional hyper-sphere of unit radius (see, e.g., Ref.~\cite{ref:levyleblond}).

The asymptotic approximation to the velocity $\velv(\sigo)$ of the vortex has a simple, {\it geometrical\/} structure to it.  This structure becomes evident via the introduction, at each point $\Sig$ of the manifold, of a pair of vectors $\shop^{1}(\Sig)$ and $\shop^{2}(\Sig)$ that are normalized, orthogonal to one another, and orthogonal to the tangent plane to the vortex sub-manifold at $\dfc(\Sig)$,
as made after Eq.~(\ref{FIG:firstNs}).  The vectors $\shop^{1}$ and $\shop^{2}$ span the manifold of dimension two complementary to vortex sub-manifold, playing the role played by the normal vector of a hyper-surface (i.e., a sub-manifold of codimension 1).
The choice of the vectors $\shop^{1}(\Sig)$ and $\shop^{2}(\Sig)$ is not unique; alternative linear combinations of them, provided the pair remains orthonormal to one another and orthogonal to the tangent plane, will do equally well.  However, what {\it is\/} independent of this choice, is unique and, from a geometric perspective, is the natural object, is the skew-symmetric product of $\shop^{1}(\Sig)$ and $\shop^{2}(\Sig)$, viz., the two-form
\begin{equation}
\shop:=\shop^{1}\wedge\shop^{2},
\end{equation}
the components of which are given by
\begin{equation}
\shop_{d_{1}d_{2}}:=
\shop^{1}_{d_{1}}\,\shop^{2}_{d_{2}}-
\shop^{1}_{d_{2}}\,\shop^{2}_{d_{1}}\,,
\end{equation}
already featuring in Eq.~(\ref{FIG:SS-Ns}) et seq.  As we show in \ref{ap:shape-operators}, the right-hand side of Eq.~(\ref{eq:vel-local-approx}) can be expressed in terms of the components of the $\shop^{1}\wedge\shop^{2}$ two-form, which then reads
\begin{equation}
\velv_{d}(\sigo)\approx
\frac{\hbar}{2M}\,\metinv_{\inin \inin^{\prime}}\,
\shop_{d d^{\prime}}\,
(\newpart_{\inin}\newpart_{\inin^{\prime}}\dfc_{d^{\prime}})\,
\ln(L/\vcs).
\label{eq:vel-local-sho}
\end{equation}
Note that we have restored to dimensional form of the velocity in Eq.~(\ref{eq:vel-local-sho}), according to the prescription given in the footnote~\footnote{To restore the physical units, we multiply
velocities by $\hbar/M$,
energies by $\mdc \hbar^2/M^2$,
momenta by $\mdc \hbar/M$ and
frequencies by $\hbar/M$.
\label{ftnote}}.

Observe the role played by the geometry here.  The local shape of the vortex enters through the {\it{intrinsic}} geometric characteristic, the metric $\met$ [i.e., the first fundamental form, Eq.~(\ref{eq:def-ind-metric})], as well as through the {\it extrinsic\/} geometric characteristic
$\shop_{dd^{\prime}}\,\newpart_{\inin}\newpart_{\inin^{\prime}}\dfc_{d^{\prime}}$
(i.e., the second fundamental form \cite{ref:fundamental-forms}).
For the familiar case of $D=3$, one choice for the two vectors $\shop^1$ and $\shop^2$ is the normal and the binormal to the vortex curve.
Together with the tangent to the $1$-dimensional vortex line they form an orthonormal basis (i.e., a Frenet-Serret frame), which reflects the extrinsic geometry of the vortex line.  In contrast with its higher-dimensional extensions, the vortex line does not possess any intrinsic geometry.

\subsection{\label{sec:vel_sp_cases}
Velocity of circular, spherical and hyper-spherical vortices: exact and asymptotic results}

We now examine the velocity of vortices for the illustrative cases of circular, spherical and hyper-spherical $\Dsl$-dimensional vortices of radius $L$, embedded in $D$-dimensional space.  In such cases, the vortex sub-manifolds can be taken to have the following form:
\begin{eqnarray}
\dfc(\sig)&=&
\,L(
\sin\sig_{1}\sin\sig_{2}\cdots\sin\sig_{\Dsl-1}\sin\sig_{\Dsl},
\nonumber\\&&\phantom{L(}
\sin\sig_{1}\sin\sig_{2}\cdots\sin\sig_{\Dsl-1}\cos\sig_{\Dsl},
\nonumber\\&&\phantom{L(}
\sin\sig_{1}\sin\sig_{2}\cdots\cos\sig_{\Dsl-1},
\nonumber\\&&\phantom{L(}
\ldots\ldots\ldots\ldots\ldots,\,
\nonumber\\&&\phantom{L(}
\sin \sig_1 \cos \sig_2,
\nonumber\\&&\phantom{L(}
\cos\sig_{1},
\nonumber\\&&\phantom{L(}
0),
\label{eq:param-hypersph}
\end{eqnarray}
with the $\Dsl$ parameters
$(\sig_{1},\ldots,\sig_{\Dsl})$
respectively ranging over intervals of length
$(\pi,2\pi,\ldots,2\pi)$,
and where, for each $D$, we have chosen the manifold to be centered at the coordinate origin and oriented so that it is confined to the $\pos_{D}=0$ hyper-plane.
For example, in the case of $D=3$, the vortex sub-manifold is then a ring of radius $L$, lying in the $\pos_{1}$-$\pos_{2}$ plane.
As the vortices are circular, spherical or hyper-spherical, rotational invariance ensures that all points on them move with a common velocity, and that this velocity points along the direction perpendicular to the $(D-1)$-dimensional plane in which the vortices are confined, i.e., the $D$ direction.

Formally, we can write an exact expression for the velocity $\velv(\sigo)$ of any point $\sigo$ on the manifold, by substituting $\dfc(\sig)$ into Eq.~(\ref{eq:vel-local}).
Thus, may choose any point on the manifold at which to compute the velocity, and for convenience we choose the point given by $\sig_1=0$ [i.e., $\dfc=(0,\ldots,0, L, 0)$].
Two convenient vectors orthogonal to each other and normal to the vortex sub-manifold there are:
\begin{equation}
\shop^{1}=(0,0,\ldots,0,0,1)\quad
{\rm{ and }}\quad
\shop^{2}=(0,0,\ldots,0,1,0).
\end{equation}
By substituting $\dfc$, $\shop^{1}$ and $\shop^{2}$ into Eq.~(\ref{eq:vel-local-sho}) we obtain
\begin{equation}
\velv_{D}(\sigo)=
\frac{1}{2L}
\ln\left({L}/{\vcs}\right),
\label{eq:vel-gen-D}
\end{equation}
regardless of the dimension $D$.

\subsection{\label{sec:oscillations}
Motion of weak distortions of linear, planar and hyper-planar vortices}

\begin{figure}[htbp]
    \centering
    \includegraphics[width=.25\textwidth]{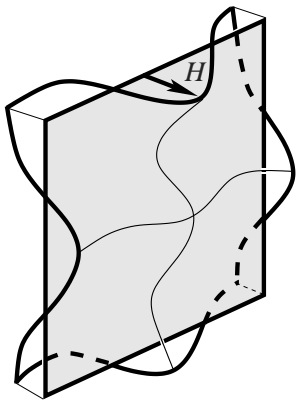}
	\caption{A weak distortion of a hyper-planar vortex in a four-dimensional superfluid.
    The height function $\height$ characterizes the distortion, which is perpendicular to the
    undistorted plane.}
\label{FIG:2d-oscillations}
\end{figure}

Next, we assume that the manifold is a weak distortion of a $\Dsl$-dimensional hyperplane, and examine the motion of this distortion that results from it being transported by the flow.
Such hyper-planar vortices are the generalization to the $D$-dimensional setting of the one-dimensional straight-line vortex in the three-dimensional setting---first studied for normal fluids by Thomson in 1880~\cite{ref:KelvinVibrations}---and the flat plane vortex in the four-dimensional setting.
The points on the undistorted hyper-plane are given by
\begin{equation}
x=\flat(\sig),
\end{equation}
where $\flat(s)=\sum_{\inin=1}^{\Dsl}\sig_{\inin}\,e_{\inin}$, the set $\{e_{d}\}_{d=1}^{D}$ is a complete orthonormal Euclidian basis in the ambient space,
and the variables $\sig$ range without bound.
To obtain a weak distortion of the flat manifold we augment $\flat(\sig)$ with a \lq\lq height\rq\rq\ function,
\begin{equation}
\height(\sig)=\height_{D-1}(\sig)\,\bve_{D-1}+\height_{D}(\sig)\,\bve_{D},
\end{equation}
which has nonzero components
$\height_{D-1}$ and $\height_{D}$ only,
i.e., it points in the two directions transverse to the flat manifold.
Figure~\ref{FIG:2d-oscillations} illustrates the weak distortion of a plane in $D=4$.  Thus, the points on the weak distortion of the flat manifold are given by
\begin{equation}
x=\flat(\sig)+\height(\sig).
\end{equation}

We now analyze the dynamics of the small perturbations (i.e., ripples) that propagate on the manifold, by constructing an equation of motion for them.  The first ingredient is the velocity $\velv$ of any point $\sigoM$ on the manifold, which we determine by applying Eq.~(\ref{eq:vel-local}), subject to the replacement
\begin{equation}
\dfc(\sig)\rightarrow\flat(\sig)+\height(\sig).
\label{eq:height-function}
\end{equation}
Thus, we arrive at the flow velocity at the point $\pos=\dfc(\sigoM)$:
\begin{eqnarray}
\velv_{d_{D}}(\sigoM)
&=&
2\pi \Dsl \frac{\Gamma(\Dsl/2)}{4\pi^{D/2}}
\epsilon_{d_{1}\cdots d_{D}}
\int d^{\Dsl}\sig\, \varepsilon_{\inin_1 \cdots \inin_{\Dsl}}
\nonumber
\\
\noalign{\smallskip}
&&\qquad\quad\times
\newpart_{\inin_{1}}(\flat_{d_{1}}(\sig)+\height_{d_{1}}(\sig))
\cdots
\newpart_{\inin_{\Dsl}}(\flat_{d_{\Dsl}}(\sig)+\height_{d_{\Dsl}}(\sig))
\nonumber
\\
\noalign{\smallskip}
&&\qquad\qquad\qquad\times
\frac{\big(\flat(\sigoM)-\flat(\sig)+\height(\sigoM)-\height(\sig)\big)_{d_{D-1}}}
{\big\vert{\flat(\sigoM)-\flat(\sig)+\height(\sigoM)-\height(\sig)}\big\vert^{D}}.
\label{eq:nearlyflat}
\end{eqnarray}
We are concerned with small distortions, and we thus expand the right hand side of Eq.~(\ref{eq:nearlyflat}) to first order in the height $\height$; the details of this expansion are given in \ref{ap:exp-brane-vel}.
For all points $\sigoM$ of the manifold, the resulting velocity $\velv(\sigoM)$ points in a direction transverse to the flat manifold, its only nonzero components being approximately given by
\begin{eqnarray}
\velv_{\gamma^{\prime}}(\sigoM)
&\approx&
2\pi\Dsl\frac{\,\Gamma(\Dsl/2)}{4\pi^{D/2}}
\epsilon_{\gamma \gamma^{\prime}}
\int\frac{d^{\Dsl}\sig}{\big\vert{\flat(\sigoM)-\flat(\sig)}\big\vert^{D}}
\nonumber
\\
&&\qquad\quad
\times
\bigg\{
\Big(\height(\sigoM)-\height(\sig)\Big)_{\gamma}-
\sum_{\nu=1}^{\Dsl}
(\sigoM-\sig)_\nu
\frac{\partial\height_{\gamma}(\sig)}{\partial\sig_{\nu}}
\bigg\},
\label{eq:vel-first-orderH}
\end{eqnarray}
where the indices $\gamma$ and $\gamma^{\prime}$
(which, are summed over when repeated)
take only the values $(D-1)$ and $D$, and $\epsilon_{\gamma \gamma^{\prime}}$ is the skew-symmetric tensor with $\epsilon_{D-1,D}:=+1$

We pause to note that the superflow velocity field $\vel(\pos)$  is translationally covariant, in the sense that a global shift in the position of the vortex sub-manifold by a constant vector $C$ (i.e., $\dfc\rightarrow\dfc+C$) produces the shift of the superflow velocity field $\vel(\pos)\rightarrow\vel(\pos-C)$.
The vortex velocity given by Eq.~(\ref{eq:nearlyflat}), and its approximation Eq.~(\ref{eq:vel-first-orderH}), both transform in accordance with this requirement.
In particular, it should be noted that Eq.~(\ref{eq:vel-first-orderH}) is invariant under translations of the distortion of the form $\height_\gamma(\sigoM)=\height_\gamma(\sigoM)+C_\gamma$.

Returning to the calculation of the vortex velocity for the case of weak distortions from hyper-planarity, the form of the denominator in the integral in Eq.~(\ref{eq:vel-first-orderH}) suggests that the dominant contribution to $\velv_{\gamma^{\prime}}(\sigo)$ comes from short distances, i.e., $\sig\approx\sigoM$.  We therefore approximate this integral by making a Taylor expansion of the term $\height(\sigoM)$ about $\sigoM=\sig$ to second order, thus arriving at the approximate vortex velocity
\begin{eqnarray}
\velv_{\gamma^{\prime}}(\sigoM)
&\approx&
2\pi\Dsl \,\frac{\Gamma(\Dsl/2)}{4\pi^{D/2}}
\,\epsilon_{\gamma \gamma^{\prime}}
\int\frac{d^{\Dsl}\sig}
{\big( \sum_{\nu^{\prime}}(\sig-\sigoM)_{\nu^{\prime}}^2 \big)^{D/2}}
\nonumber
\\
&&\qquad\qquad\qquad\qquad\qquad\times
(\sigoM-\sig)_{\nu} (\sigoM-\sig)_{\bar{\nu}}\,\frac{1}{2}\,
\frac{\partial^2 \height_{\gamma}}{\partial \sig_{\nu} \partial \sig_{\bar{\nu}}}.
\label{eq:vel-dominant}
\end{eqnarray}

To obtain a closed system of equations for the dynamics of the height function, we observe that the superflow carries the vortex sub-manifold with it, so that the evolution in time $t$ of the height is given by
\begin{equation}
\partial\height_{\gamma}(\sigoM,t)/\partial t=\velv_{\gamma}(\sigoM,t).
\label{eq:EOM-approx}
\end{equation}
In the approximation~(\ref{eq:vel-dominant}), this equation of motion becomes a first-order differential (in time),
nonlocal (in \lq\lq  internal space\rlap,\rq\rq\ i.e., in the coordinates $\sigoM$ that span the undistorted, hyper-planar vortex sub-manifold) system of two coupled linear equations that is translationally invariant in both (internal) space and time.  As such, it is reducible to a two-by-two matrix algebraic form under Fourier transformation in both (internal) space and time, which we define as follows:
\begin{eqnarray}
\widehat{L}(\sigoMhat,\that)
&=&
\int{d^{\Dsl}\sigoM}\,\,
{\rm e}^{-iq\cdot\sigoM}
\int{dt}\,\,
{\rm e}^{i\omega t}\,
L(\sigoM,t),
\\
L(\sigoM,t)
&=&
\int\frac{d^{\Dsl}\sigoMhat}{(2\pi)^{\Dsl}}\,\,
{\rm e}^{iq\cdot\sigoM}
\int\frac{d\that}{2\pi}\,\,
{\rm e}^{-i\omega t}\,
\widehat{L}(\sigoMhat,\that).
\label{eq:Fourier-defs}
\end{eqnarray}
Computing the spatial Fourier transform of Eq.~(\ref{eq:EOM-approx}) is addressed in \ref{ap:four-an-brane}; the temporal Fourier transform is elementary.  Introducing, for convenience, the coefficient
\begin{equation}
\myalpha(\vert{k}\vert):=
\frac{\pi^{\Dsl/2}}{\Gamma(\Dsl/2)}\,
k^{2}\ln\left(1/\vert{k}\vert\xi\right),
\end{equation}
arrived at via the spatial Fourier transform (see \ref{ap:four-an-brane}),
the equation of motion becomes
\begin{equation}
-i\omega
\widehat{\height}_{\gamma}(\sigoMhat,\that)=
\myalpha(\vert{q}\vert)\,\epsilon_{\gamma \gamma^{\prime}}\,
\widehat{\height}_{\gamma^{\prime}}(\sigoMhat,\that).
\label{eq:EOM-algebraic}
\end{equation}
The solvability condition for this system is
\begin{equation}
\det\left(
  \begin{array}{ c c }
     i\omega                     & \myalpha(\vert{q}\vert) \\
   -\myalpha(\vert{q}\vert)   & i\omega
  \end{array} \right)=0,
\end{equation}
and gives rise to the following dispersion relation for the spectrum of normal modes of oscillation, with two modes for each value of the $\Dsl$-vector $q$ conjugate to the position in the flat vortex sub-manifold hyper-plane:
\begin{equation}
\omega(\vert{q}\vert)=
\pm\frac{\pi^{\Dsl/2}}{\Gamma(\Dsl/2)}\,\,
\frac{\hbar}{M} \,\vert{q}\vert^{2}
\ln\left(1/\vert{q}\vert\xi\right),
\label{eq:en-osc-modes}
\end{equation}
where we have restored the dimensional factors.
The normal modes are themselves given by
\begin{equation}
\left(
  \begin{array}{ c }
     \widehat{\height}_{D-1}^{\pm} \\
     \widehat{\height}_{D\phantom{-1}}^{\pm}
  \end{array}
\right)
\approx
\left(
  \begin{array}{ c }
     1 \\
     \mp i
  \end{array} \right)
  e^{i(q \sigoM \mp \omega t)},
\end{equation}
and these correspond to left and right circularly polarized oscillations.

\section{%
\label{sec:energcalc}
Superflow kinetic energy for given vorticial content}

\subsection{%
\label{sec:energy}
Exact evaluation}
We now use the solution for the velocity field, Eq.~(\ref{eq:velfieldsource}), together with the equation for the kinetic energy $E$ of the velocity field, Eq.~(\ref{eq:KEintegral}), to obtain the kinetic energy in terms of the source $\source$:
\begin{eqnarray}
E
&=&
\frac{1}{2}\int_{\vol} d^{D}\!\pos\,\vel_{d}(\pos)\,\vel_{d}(\pos)
=
\frac{1}{2}\int_{\vol} d^{D}\!\pos\,\vel_{d}^{\ast}(\pos)\,\vel_{d}^{\phantom{\ast}}(\pos)
\nonumber\\
&=&
\frac{1}{2}\delta_{d_{D}^{\prime}d_{D}}
\left(\frac{2\pi}{\Dsl !}\right)^{2}
\int_{\vol} d^{D}\!\pos\,
\epsilon_{d_{1}^{\prime}\cdots d_{D}^{\prime}}
\int\dbar^{D}\!q^{\prime}\,
q_{d_{D-1}^{\prime}}^{\prime}\,
{\rm e}^{-iq^{\prime}\cdot\pos}\,
\frac{\source_{d_{1}^{\prime}\cdots d_{\Dsl}^{\prime}}^{\ast}(q^{\prime})}{q^{\prime}\cdot q^{\prime}}\,
\nonumber\\&&\qquad\qquad\qquad\qquad\quad\times
\epsilon_{d_{1}\cdots d_{D}}
\int\dbar^{D}\!q\,
q_{d_{D-1}}\,{\rm e}^{iq\cdot\pos}\,
\frac{\source_{d_{1}\cdots d_{\Dsl}}(q)}{q\cdot q}
\nonumber
\\
&=&
\frac{1}{2}\delta_{d_{D}^{\prime}d_{D}}
\left(\frac{2\pi}{\Dsl !}\right)^{2}
\epsilon_{d_{1}^{\prime}\cdots d_{D}^{\prime}}
\int\dbar^{D}\!q^{\prime}\,
q_{d_{D-1}^{\prime}}^{\prime}\,
\frac{\source_{d_{1}^{\prime}\cdots d_{\Dsl}^{\prime}}^{\ast}(q^{\prime})}{q^{\prime}\cdot q^{\prime}}\,
\nonumber\\&&\qquad\qquad\qquad\times
\epsilon_{d_{1}\cdots d_{D}}
\int\dbar^{D}\!q\,
q_{d_{D-1}}\,
\frac{\source_{d_{1}\cdots d_{\Dsl}}(q)}{q\cdot q}\,
\diracbar^{(D)}\!(q^{\prime}-q)\,
\nonumber\\
&=&
\frac{1}{2}\delta_{d_{D}^{\prime}d_{D}}
\left(\frac{2\pi}{\Dsl !}\right)^{2}
\epsilon_{d_{1}^{\prime}\cdots d_{D}^{\prime}}
\epsilon_{d_{1}\cdots d_{D}}
\nonumber\\&&\qquad\qquad\qquad\times
\int\dbar^{D}q\,
\source_{d_{1}^{\prime}\cdots d_{\Dsl}^{\prime}}^{\ast}(q)\,
\source_{d_{1}\cdots d_{\Dsl}}(q)\,
\frac{q_{d_{D-1}^{\prime}}\,q_{d_{D-1}}}{(q\cdot q)^{2}},
\end{eqnarray}
where $\diracbar^{(D)}\!(q):=(2\pi)^{D}\dirac^{(D)}\!(q)$.  For a source $\source$ given by a single vortex, as in Eq.~(\ref{eq:current-allD}), we then have the more explicit form for the energy:
\begin{eqnarray}
&&E=
\frac{\delta_{d_{D}^{\prime}d_{D}}}{2}
\left(\frac{2\pi}{\Dsl !}\right)^{2}
\epsilon_{d_{1}^{\prime}\cdots d_{D}^{\prime}}
\epsilon_{d_{1}\cdots d_{D}}
\int d^{\Dsl}\sig^{\prime}\,
\varepsilon_{\inin_{1}^{\prime}\cdots\inin_{\Dsl}^{\prime}}
\big(\newpart_{\inin_{1}^{\prime}}\dfc_{d_{1}^{\prime}}(\sig^{\prime})\big)
\cdots
\big(\newpart_{\inin_{\Dsl}^{\prime}}\dfc_{d_{\Dsl}^{\prime}}(\sig^{\prime})\big)
\nonumber\\&&\qquad\qquad\qquad\times
\int d^{\Dsl}\sig\,
\varepsilon_{\inin_{1}\cdots\inin_{\Dsl}}
\big(\newpart_{\inin_{1}}\dfc_{d_{1}}(\sig)\big)
\cdots
\big(\newpart_{\inin_{\Dsl}}\dfc_{d_{\Dsl}}(\sig)\big)
\nonumber\\&&\qquad\qquad\qquad\qquad\qquad\qquad\times
\int\dbar^{D}q\,
{\rm e}^{iq\cdot\left(\dfc(\sig^{\prime})-\dfc(\sig)\right)}
\frac{q_{d_{D-1}^{\prime}}\,q_{d_{D-1}}}{(q\cdot q)^{2}}.
\label{eq:q-int-for-energy}
\end{eqnarray}
Writing $\sep$ for $\left(\dfc(\sig^{\prime})-\dfc(\sig)\right)$ and using the result for the $q$ integration given in \ref{ap:invariant-int}, the energy becomes
\begin{eqnarray}
&&E=
\frac{1}{2}
\left(\frac{2\pi}{\Dsl !}\right)^{2}
\epsilon_{d_{1}^{\prime}\cdots d_{D}^{\prime}}
\epsilon_{d_{1}\cdots d_{D}}
\delta_{d_{D}^{\prime}d_{D}}
\int d^{\Dsl}\sig^{\prime}\,
\varepsilon_{\inin_{1}^{\prime}\cdots\inin_{\Dsl}^{\prime}}
\big(\newpart_{\inin_{1}^{\prime}}\dfc_{d_{1}^{\prime}}(\sig^{\prime})\big)
\cdots
\big(\newpart_{\inin_{\Dsl}^{\prime}}\dfc_{d_{\Dsl}^{\prime}}(\sig^{\prime})\big)
\nonumber\\&&\qquad\qquad\times
\int d^{\Dsl}\sig\,
\varepsilon_{\inin_{1}\cdots\inin_{\Dsl}}
\big(\newpart_{\inin_{1}}\dfc_{d_{1}}(\sig)\big)
\cdots
\big(\newpart_{\inin_{\Dsl}}\dfc_{d_{\Dsl}}(\sig)\big)
\nonumber\\&&\qquad\qquad\qquad\qquad\times
\frac{1}{\left(\sep\cdot\sep\right)^{\Dsl/2}}
\left(\delta_{d_{D-1}^{\prime}d_{D-1}}\,{\cal P}_{1}+
\frac{\sep_{d_{D-1}^{\prime}}\,\sep_{d_{D-1}}}{\sep\cdot\sep}\,{\cal P}_{2}\right),
\label{eq:energ-exact}
\end{eqnarray}
where ${\cal P}_{1}$ and ${\cal P}_{2}$ depend only on $D$ and are given in \ref{ap:invariant-int}.

The final step in constructing the energy of a single vortex involves contracting the $D$-dimensional $\epsilon$ symbols and taking advantage of the skew-symmetry of the sources.  Thus, we arrive at the result
\begin{eqnarray}
&&E
=
\frac{1}{2}
\left(\frac{2\pi}{\Dsl !}\right)^{2}
\frac{1}{(2\pi)^{D}}
\surf_{D-1}\,\sqrt{\pi}\,\Gamma\left(\textstyle{\frac{D}{2}-\frac{1}{2}}\right)\,
2^{\Dsl-2}\,\Gamma\left(\textstyle{\frac{D}{2}-1}\right)
\nonumber\\&&\qquad\times
\int d^{\Dsl}\sig^{\prime}\,
\varepsilon_{\inin_{1}^{\prime}\cdots\inin_{\Dsl}^{\prime}}
\big(\newpart_{\inin_{1}^{\prime}}\dfc_{d_{1}}(\sig^{\prime})\big)
\cdots
\big(\newpart_{\inin_{\Dsl-1}^{\prime}}\dfc_{d_{\Dsl-1}}(\sig^{\prime})\big)
\big(\newpart_{\inin_{\Dsl}^{\prime}}\dfc_{d_{\Dsl}^{\prime}}(\sig^{\prime})\big)
\nonumber\\&&\qquad\qquad\times
\int d^{\Dsl}\sig\,
\varepsilon_{\inin_{1}\cdots\inin_{\Dsl}}
\big(\newpart_{\inin_{1}}\dfc_{d_{1}}(\sig)\big)
\cdots
\big(\newpart_{\inin_{\Dsl-1}}\dfc_{d_{\Dsl-1}}(\sig)\big)
\big(\newpart_{\inin_{\Dsl}}\dfc_{d_{\Dsl}}(\sig)\big)
\nonumber\\&&\qquad\qquad\qquad\times
\frac{1}{\left(\sep\cdot\sep\right)^{\Dsl/2}}
\Big\{
(4-D)\Dsl !\,\delta_{d_{\Dsl}^{\prime}d_{\Dsl}}
+\Dsl^{2}\Dsl !\,
\frac{\sep_{d_{\Dsl}^{\prime}}\,\sep_{d_{\Dsl}}}{\sep\cdot\sep}
\Big\}.
\label{eq:double}
\end{eqnarray}
The factor $\surf_{D-1}$ is the surface area of a $(D-1)$-dimensional hyper-sphere of unit radius (see, e.g., Ref.~\cite{ref:levyleblond}), and $\Gamma(z)$ is the standard Gamma function (see, e.g., Ref.~\cite{ref:Abram-handbook}).

Formula~(\ref{eq:double}) gives the kinetic energy in terms of the local geometry of \lq\lq patches\rq\rq\thinspace\ of the vortex sub-manifold and interactions between such \lq\lq patches\rlap.\rq\rq\thinspace\ The asymptotically dominant contributions to the energy come from nearby patches, and we evaluate these contributions in the following section.  The linearity of the theory ensures that the energy associated with configurations involving multiple, rather than just one, vortices is obtained via an elementary extension, comprising self-interactions terms, such as the one just given, for each vortex, as well as mutual interaction terms for pairs of distinct vortices.

\subsection{\label{sec:asymptotics}
Asymptotics and geometry}

We now analyze the energy of a vortex asymptotically, subject to the conditions discussed earlier in Section~\ref{sec:velocity}, (i.e., the characteristic linear dimensions and radii of curvature of the vortex are much larger than the short-distance cut-off $\vcs$).  In this regime, the dominant contribution to the energy is associated with the large values of the integrand that occur when the ($\Dsl$-dimensional) integration variables $\sig$ and $\sig^{\prime}$ are nearly coincident, so that the denominator $(\sep\cdot\sep)^{\Dsl/2}$ is small.  To collect such contributions, it is useful to change variables from $(\sig,\sig^{\prime})$ to $(\Sig,\sigrel)$ as follows:
\begin{eqnarray}
\sig_{\inin}&:=&\Sig_{\inin},
\\
\sig_{\inin}^{\prime}&:=&\Sig_{\inin}+\sigrel_{\inin},
\end{eqnarray}
the Jacobian for this transformation being unity,
\begin{equation}
\frac{\partial(\sig,\sig^{\prime})}{\partial(\Sig,\sigrel)}
=\left[\text{det}\pmatrix{1&1\cr 0&1}\right]^{\Dsl}=1,
\end{equation}
so that
$\int d^{\Dsl}\!\sig^{\prime}\int d^{\Dsl}\!\sig\ldots=
\int d^{\Dsl}\!\Sig\int d^{\Dsl}\!\sigrel\ldots\,\,$.

The energy formula~(\ref{eq:double}) comprises two pieces, and we now consider them in turn.
The first, (a), involves the integral
\begin{eqnarray}
&&
\int d^{\Dsl}\Sig\,
\int \frac{d^{\Dsl}\!\sigrel}
{\left(\sep\cdot\sep\right)^{\Dsl/2}}
\,\varepsilon_{\inin_{1}^{\prime}\cdots\inin_{\Dsl}^{\prime}}\,
\big(\newpart_{\inin_{1}^{\prime}}\dfc_{d_{1}}(\sigo-\sig)\big)
\cdots
\big(\newpart_{\inin_{\Dsl}^{\prime}}\dfc_{d_{\Dsl}}(\sigo-\sig)\big)
\nonumber\\
&&\qquad\qquad\qquad\qquad\qquad\times
\varepsilon_{\inin_{1}\cdots\inin_{\Dsl}}
\big(\newpart_{\inin_{1}}\dfc_{d_{1}}(\sig)\big)
\cdots
\big(\newpart_{\inin_{\Dsl}}\dfc_{d_{\Dsl}}(\sig)\big).
\label{eq:double-A}
\end{eqnarray}
The second, (b), involves the integral
\begin{eqnarray}
&&
\int d^{\Dsl}\Sig\,
\int d^{\Dsl}\!\sigrel\,
\frac{\sep_{d_{\Dsl}^{\prime}}\,\sep_{d_{\Dsl}}}
{\left(\sep\cdot\sep\right)^{D/2}}\,
\nonumber\\&&\qquad\times
\varepsilon_{\inin_{1}^{\prime}\cdots\inin_{\Dsl}^{\prime}}\,
\big(\newpart_{\inin_{1}^{\prime}}\dfc_{d_{1}}(\sigo-\sig)\big)
\cdots
\big(\newpart_{\inin_{\Dsl-1}^{\prime}}\dfc_{d_{\Dsl-1}}(\sigo-\sig)\big)
\big(\newpart_{\inin_{\Dsl}^{\prime}}\dfc_{d_{\Dsl}^{\prime}}(\sigo-\sig)\big)
  \nonumber\\&&\qquad\times
\varepsilon_{\inin_{1}\cdots\inin_{\Dsl}}
\big(\newpart_{\inin_{1}}\dfc_{d_{1}}(\sig)\big)
\cdots
\big(\newpart_{\inin_{\Dsl-1}}\dfc_{d_{\Dsl-1}}(\sig)\big)
\big(\newpart_{\inin_{\Dsl}}\dfc_{d_{\Dsl}}(\sig)\big).
\label{eq:double-B}
\end{eqnarray}
Our strategy will be to Taylor-expand the integrands in terms~(a) and (b) about the point $\sigrel=0$, keeping only the leading-order terms.

Let us focus on term~(a).  It is adequate to approximate the numerator at zeroth order:
\[
\varepsilon_{\inin_{1}^{\prime}\cdots\inin_{\Dsl}^{\prime}}
\big(\newpart_{\inin_{1}^{\prime}}\dfc_{d_{1}}(\Sig)\big)
\!\cdots\!
\big(\newpart_{\inin_{\Dsl}^{\prime}}\dfc_{d_{\Dsl}}(\Sig)\big)
\,\varepsilon_{\inin_{1}\cdots\inin_{\Dsl}}\,
\big(\newpart_{\inin_{1}}\dfc_{d_{1}}(\Sig)\big)
\!\cdots\!
\big(\newpart_{\inin_{\Dsl}}\dfc_{d_{\Dsl}}(\Sig)\big)\,.
\]
Note the occurrence of $\Dsl$ factors of the induced metric $\met_{\inin \inin^{'}}$, defined in Eq.~(\ref{eq:def-ind-metric}).
In terms of $\met$, the numerator of term~(a) becomes
\begin{equation}
\varepsilon_{\inin_{1}^{\prime}\cdots\inin_{\Dsl}^{\prime}}\,
\varepsilon_{\inin_{1}\cdots\inin_{\Dsl}}\,
\met_{\inin_{1}^{\prime}\inin_{1}}(\Sig)\,
\met_{\inin_{2}^{\prime}\inin_{2}}(\Sig)\cdots
\met_{\inin_{\Dsl}^{\prime}\inin_{\Dsl}}(\Sig),
\label{eq:double-A-num-met}
\end{equation}
and, by the well-known formula for determinants, this is
\begin{equation}
\Dsl !\,\,\text{det}\,\met\,.
\label{eq:double-A-num-det}
\end{equation}
To complete the approximate computation of term~(a) we examine the denominator,
\begin{equation}
\left(\sep\cdot\sep\right)^{-\Dsl/2}=
\vert\dfc(\Sig+\sigrel)-\dfc(\Sig)\vert^{-\Dsl}.
\end{equation}
By Taylor-expanding for small $\sigrel$ and retaining only the leading-order contribution, we find that the denominator becomes
\begin{equation}
\left(\sigrel_{\inin^{\prime}}\,\met_{\inin^{\prime}\inin}\,\sigrel_{\inin}\right)^{-\Dsl/2}.
\end{equation}
Thus, for term~(a) we have the approximate result
\begin{equation}
\int d^{\Dsl}\Sig\int d^{\Dsl}\sigrel\,
\frac{\Dsl !\,\,{\text det}\,\met(\Sig)}
{\left(\sigrel_{\inin^{\prime}}\,\met_{\inin^{\prime}\inin}(\Sig)\,
\sigrel_{\inin}\right)^{\Dsl/2}}\,.
\end{equation}
A convenient way to analyze the integration over $\sigrel$ is to make a $\Dsl$-dimensional orthogonal transformation from $\sigrel$ to coordinates $\prinrel$ that diagonalize $\met$, the eigenvalues of which we denote by $\{\metev_{\inin}\}_{\inin=1}^{\Dsl}$, and then to \lq\lq squash\rq\rq\ these coordinates to $\prinsq$ (defined via $\prinsq_{\inin}:=\sqrt{\metev_{\inin}}\,\prinrel_{\inin}$), so as to \lq\lq isotropize\rq\rq\ the quadratic form in the denominator.  Following these steps, postponing to the following paragraph a discussion of the integration limits, and omitting an overall factor of $\Dsl !$, our approximation for term~(a) becomes
\begin{eqnarray}
&&\int d^{\Dsl}\Sig\,\,{\text det}\,\met(\Sig)
\int\frac{d^{\Dsl}\sigrel}
{\left(\sigrel_{\inin^{\prime}}\,\met_{\inin^{\prime}\inin}(\Sig)\,
\sigrel_{\inin}\right)^{\Dsl/2}}
\nonumber\\
&&\qquad\qquad=
\int d^{\Dsl}\Sig\,\,{\text det}\,\met(\Sig)
\int\frac{d^{\Dsl}\prinrel}
{\Big[\sum_{\inin=1}^{\Dsl}
\metev_{\inin}(\Sig)\,\prinrel_{\inin}^{2}\,\Big]^{\Dsl/2}}
\nonumber\\
&&\qquad\qquad=
\int d^{\Dsl}\Sig\,\,
\frac{{\text det}\,\met(\Sig)}{\prod_{\inin=1}^{\Dsl}\sqrt{\metev_{\inin}(\Sig)}}
\int\frac{d^{\Dsl}\prinsq}
{\Big[\sum_{\inin=1}^{\Dsl}\prinsq_{\inin}^{2}\,\Big]^{\Dsl/2}}
\nonumber\\
&&\qquad\qquad=
\int d^{\Dsl}\Sig\,\sqrt{{\text det}\,\met(\Sig)}\,
\surf_{\Dsl}\int
\frac{d\prinsqmag\,\prinsqmag^{\Dsl-1}}{\prinsqmag^{\Dsl}},
\nonumber
\end{eqnarray}
where $\prinsqmag:=\sqrt{\sum_{\inin=1}^{\Dsl}\prinsq_{\inin}^{2}}$.

Up to now, we have been vague about the limits on the integrations over the $\sigrel$ variables.  In the short-distance regime there is, for each value of $\Sig$, a physically motivated cutoff that eliminates contributions from regions of $\sigrel$ in which the separation $\vert\dfc(\Sig+\sigrel)-\dfc(\Sig)\vert$ is of order $\vcs$ or smaller.  This cutoff is associated with the fact that our overall description is not intended (and is indeed unable) to capture physics on length-scales associated with the linear dimension $\vcs$ of the core of the vortices.  In the long-distance regime, there are integration limits associated with the finite size of the vortex, which we take to be of order $L$.  (Thus, we rule out of consideration vortices whose shape is strongly anisotropic; already ruled out are, e.g., rough vortices, for which the Taylor expansion of the functions describing the shapes of the vortices would be unjustified.)\thinspace\  In particular, the last step in our approximate computation of term~(a), in which we transformed to the single radial coordinate in the $\Dsl$-dimensional flat plane tangent to the vortex at the point $\sig=\Sig$, is only valid if the limits of integration possess the rotational symmetry of this tangent plane, which they typically do not.  (The case of $D=3$ is an exception.)\thinspace\  However, to within the logarithmic accuracy to which we are working, it is adequate to ignore these complications and approximate what has become the $\prinsqmag$ integral by $\ln(L/\vcs)$.  Any refinement would just change the argument of the logarithm by a multiplicative numerical factor, and this would produce an additive correction to our result that is of sub-leading order.  Thus, our approximation to term~(a) becomes
\begin{equation}
\Dsl !\,
\int d^{\Dsl}\Sig\,\sqrt{\text{det}\,\met(\Sig)}\,
\surf_{\Dsl}\ln(L/\vcs).
\end{equation}
The remaining integral, $\int d^{\Dsl}\Sig\,\sqrt{\text{det}\,\met(\Sig)}$, is familiar from differential geometry, and simply gives the $\Dsl$-dimensional volume of the sub-manifold supporting the vortex, which we shall denote by $\defvol_{\Dsl}$ [and which is a functional of $\dfc(\cdot)$].  Thus, our approximation to term~(a) becomes
\begin{equation}
\Dsl !\,\surf_{\Dsl}\,\defvol_{\Dsl}\,\ln(L/\vcs).
\label{eq:exactasy}
\end{equation}

A similar procedure allows us to find an adequate approximation to term~(b), as we now show.  In contrast with term~(a), owing to the two factors of
$\sep_{d}:=\dfc_{d}(\Sig+\sigrel)-\dfc_{d}(\Sig)$,
the numerator vanishes at zeroth order in $\sigrel$, and requires Taylor expansion,
\begin{equation}
\sep_{d}\approx\sigrel_{\inin}\,\partial\dfc_{d}/\partial\Sig_{\inin},
\end{equation}
to identify the leading-order contribution to the numerator.  The remaining factors in the numerator can be evaluated at zeroth order, which then becomes
\begin{eqnarray}
&&
\varepsilon_{\inin_{1}^{\prime}\cdots\inin_{\Dsl}^{\prime}}\,
\newpart_{\inin_{1}^{\prime}}\dfc_{d_{1}}
\cdots
\newpart_{\inin_{\Dsl-1}^{\prime}}\dfc_{d_{\Dsl-1}}\,
\newpart_{\inin_{\Dsl}^{\prime}}\dfc_{d_{\Dsl}^{\prime}}\,
\varepsilon_{\inin_{1}\cdots\inin_{\Dsl}}\,
\newpart_{\inin_{1}}\dfc_{d_{1}}
\cdots
\newpart_{\inin_{\Dsl-1}}\dfc_{d_{\Dsl-1}}\,
\newpart_{\inin_{\Dsl}}\dfc_{d_{\Dsl}}
\nonumber\\&&\qquad\qquad\qquad\qquad
\times
\sigrel_{\inin_{D-1}^{\prime}}\,\newpart_{\inin_{D-1}^{\prime}}\dfc_{d_{\Dsl}^{\prime}}\,
\sigrel_{\inin_{D-1}         }\,\newpart_{\inin_{D-1}         }\dfc_{d_{\Dsl}         }
\nonumber\\&&\quad
=
\varepsilon_{\inin_{1}\cdots\inin_{\Dsl}}\,
\varepsilon_{\inin_{1}^{\prime}\cdots\inin_{\Dsl}^{\prime}}\,
\met_{\inin_{1}^{\prime}\inin_{1}}
\cdots
\met_{\inin_{D-3}^{\prime}\inin_{D-3}}\,
\met_{\inin_{\Dsl}^{\prime}\inin_{D-1}^{\prime}}\,
\sigrel_{\inin_{D-1}^{\prime}}\,
\met_{\inin_{\Dsl}\inin_{D-1}}\,
\sigrel_{\inin_{D-1}         }
\nonumber\\&&\quad
=
\varepsilon_{\inin_{1}\cdots\inin_{\Dsl-1}\inin_{\Dsl}}\,
\varepsilon_{\inin_{1}\cdots\inin_{\Dsl-1}\inin_{D-1}^{\prime}}\,
\sigrel_{\inin_{D-1}^{\prime}}\,
\met_{\inin_{\Dsl}\inin_{D-1}}\,
\sigrel_{\inin_{D-1}         }
\nonumber\\&&\quad
=
(\Dsl-1)!\,\delta_{\inin_{\Dsl}\inin_{D-1}^{\prime}}\,
\sigrel_{\inin_{D-1}^{\prime}}\,
\met_{\inin_{\Dsl}\inin_{D-1}}\,
\sigrel_{\inin_{D-1}         }
=
\Dsl !\,\,
\sigrel_{\inin^{\prime}}\,
\met_{\inin^{\prime}\inin}\,
\sigrel_{\inin}\,,
\end{eqnarray}
in which all terms involving $\dfc$ and $\met$ are evaluated at the point $\Sig$, and where we have made use of the formula for the determinant,
\begin{equation}
\varepsilon_{\inin_{1}^{\prime}\cdots\inin_{\Dsl}^{\prime}}\,
\met_{\inin_{1}^{\prime}\inin_{1}}\,
\met_{\inin_{2}^{\prime}\inin_{2}}\cdots
\met_{\inin_{\Dsl}^{\prime}\inin_{\Dsl}}
=
\varepsilon_{\inin_{1}\cdots\inin_{\Dsl}}\,{\text det}\,\met,
\end{equation}
and, once again, introduced the induced metric $\met_{\inin^{\prime}\inin}(\Sig)$ as well as a contraction
$\varepsilon_{\inin_{1}\cdots\inin_{\Dsl-1}\inin}\,
 \varepsilon_{\inin_{1}\cdots\inin_{\Dsl-1}\inin^{\prime}}
=(\Dsl-1)!\,\delta_{\inin\inin^{\prime}}$.
As for the the denominator of term~(b), this we Taylor-expand in the manner already used for term~(a), noting that the overall powers in these denominators differ by unity.  Thus, apart from an overall factor of $(\Dsl-1)!$, we have for term~(b) the approximate result
\begin{eqnarray}
&&
\int d^{\Dsl}\Sig\,{\text det}\,\met(\Sig)
\int d^{\Dsl}\sigrel\,
\frac{\sigrel_{\inin^{\prime}}\,
\met_{\inin^{\prime}\inin}(\Sig)\,
\sigrel_{\inin}}
{\left(\sigrel_{\inin^{\prime}}\met_{\inin^{\prime}\inin}(\Sig)\,
\sigrel_{\inin}\right)^{D/2}}\,
\nonumber\\&&
\qquad\qquad=
\int d^{\Dsl}\Sig\,{\text det}\,\met(\Sig)
\int\frac{d^{\Dsl}\sigrel}
{\left(\sigrel_{\inin^{\prime}}\,\met_{\inin^{\prime}\inin}(\Sig)\,
\sigrel_{\inin}\right)^{\Dsl/2}},
\end{eqnarray}
which is of precisely the structure that arose in the analysis of term~(a), and thus the approximate result for term~(b) becomes
\begin{equation}
(\Dsl-1)!\,\surf_{\Dsl}\,\defvol_{\Dsl}\,\ln(L/\vcs).
\end{equation}

Finally, we assemble terms~(a) and (b) to arrive at an asymptotic formula for the energy of the equilibrium flow in the presence of a vorticial defect, which reads
\begin{eqnarray}
E\approx
\frac{\pi^{2-D}}{4}\sqrt{\pi}\,
\Gamma\left(\textstyle{\frac{D}{2}-\frac{1}{2}}\right)
\Gamma\left(\textstyle{\frac{D}{2}-1          }\right)
\surf_{D-1}\,\surf_{\Dsl}\,
\ln(L/\vcs)\,\defvol_{\Dsl}.
\end{eqnarray}
This formula simplifies, using the equation for $\surf_{\deldef}$, Ref.~\cite{ref:levyleblond}), to give
\begin{eqnarray}
E\approx
\left(\mdc \frac{\hbar^2}{M^2}\right)\pi\,\defvol_{\Dsl}\,\ln(L/\vcs),
\label{eq:energy-asy-anyD}
\end{eqnarray}
where we have restored the dimensional factors.

\subsection{\label{sec:elementary}
Elementary argument for the energy}

In the previous subsection, \ref{sec:asymptotics}, we have shown that, in the asymptotic limit of a large, smooth, $\Dsl$-dimensional vortex sub-manifold having the topology of the surface of a $(D-1)$-dimensional hyper-sphere, the kinetic energy depends on the geometry of the sub-manifold only through its $\Dsl$-dimensional volume $\defvol_{\Dsl}$.  This result is consistent with the following elementary argument.

Imagine starting with such a vortex sub-manifold, and \lq\lq morphing\rq\rq\ its shape into that of a hyper-sphere having the same $\Dsl$-dimensional volume.  Next, cut the hyper-sphere along an equator, so that it becomes two hyper-hemispheres of equal size.  Then, distort the hyper-hemispheres into flat hyper-planes separated by $\alpha L$ (i.e., a distance of order $L$), maintaining their $\Dsl$-dimensional volumes at $\frac{1}{2}\defvol_{\Dsl}$, and imposing periodic boundary conditions on opposing sides of the hyper-planes.  This last step is bound to involve some stretching of the sub-manifolds, but let us assume that this will only affect the energy sub-dominantly.  We choose the separation of the hyper-planes to be ${\cal O}(L)$, so as to best maintain the geometry of the original sub-manifold.  We can think of the resulting $D$-dimensional system as comprising a $\Dsl$-dimensional stack of identical, two-dimensional, films of superfluid having thicknesses $\vcs$ in every one of their $\Dsl$ thin dimensions, each film containing a point-like vortex--anti-vortex pair.

Now, owing to the translational and inversion symmetry of the stack, the flow in each film is confined to the film and is simply that associated with vortices in the film, and therefore the energy of the flow is given by the formula:
$2\pi\vcs^{\Dsl}\,\ln(\alpha L/\vcs)$ (see, e.g., Ref.~\cite{ref:Thouless-book}).
Furthermore, the number of films is
$\frac{1}{2}\defvol_{\Dsl}/\vcs^{\Dsl}$.
Thus, consistent with Eq.~(\ref{eq:energy-asy-anyD}), the total energy is given by
\begin{equation}
\frac{\defvol_{\Dsl}}{2\vcs^{\Dsl}}\times 2\pi \vcs^{\Dsl}\,\ln\left(\frac{\alpha L}{\vcs}\right)=
\pi\defvol_{\Dsl}\,\left\{\ln\frac{L}{\vcs}+{\cal O}(1)\right\},
\end{equation}
where the ${\cal O}(1)$ correction is associated with the factor $\alpha$.

Sub-leading corrections to this result for the energy will not depend solely on the vortex sub-manifold volume but also on its more refined local and global geometrical characteristics.

\subsection{\label{sec:cases}
Energy of circular, spherical and hyper-spherical vortices: exact and asymptotic results }

In this section, we analyze the energy of a geometrically (and not just topologically) hyper-spherical vortex sub-manifold.  The function $\dfc(\sig)$ for such a vortex is given by Eq.~(\ref{eq:param-hypersph}), and by substituting this form into Eq.~(\ref{eq:energ-exact}) we obtain a formally exact result for the energy.

To illustrate this point, consider the case of a two-dimensional vortex in $D=4$, for which the energy reads
\begin{eqnarray}
&&E=\frac{1}{8}L^{2}
\int_{-\pi}^{\pi}     d\sig_1\, d\sig^{\prime}_1
\int_{-\pi/2}^{\pi/2} d\sig_2\, d\sig^{\prime}_2\,
\cos(\sig_2)\,
\cos(\sig^{\prime}_2)
\nonumber \\
&&\qquad\quad\times
\left\{
\frac{2}{1-\cos(\sig_1-\sig^{\prime}_1)\,\cos\sig_2\,\cos\sig^{\prime}_2
-\sin\sig_2\,\sin\sig^{\prime}_2}-1
\right\}.
\label{eq:exact-en-D=4}
\end{eqnarray}
As with the case of arbitrary $D$, the integrand in Eq.~(\ref{eq:exact-en-D=4}) is singular for
$(\sig_1,\sig_2)=(\sig^{\prime}_1,\sig^{\prime}_2)$; therefore, we follow the  procedure, described in detail in Section~\ref{sec:asymptotics}, of expanding the integrand around this point of singularity, hence obtaining the asymptotic result that the energy is given by
\begin{equation}
E\approx 4\pi^{2}\,L^{2}\,\ln(L/\vcs).
\end{equation}

Turning now to arbitrary dimension $D$, the energy of a hyper-spherical vortex can be obtained by replacing the volume of the manifold in Eq.~(\ref{eq:energy-asy-anyD}) by the surface area of a $(D-1)$-dimensional hyper-sphere, viz.,
$\surf_{D-1}L^{\Dsl}$ (see Ref.~\cite{ref:levyleblond}),
so that we arrive at
\begin{equation}
E\approx \pi\,\surf_{D-1}\,L^{\Dsl}\,\ln(L/\vcs).
\end{equation}

\subsection{\label{sec:dispersion}
Energy-momentum relation for vortices}

We now use the results for the energies and velocities of circular, spherical and hyper-spherical vortices, obtained respectively in Secs.~\ref{sec:vel_sp_cases} and \ref{sec:cases}, to construct the momentum $\mom$ of such vortices and, hence, their energy-momentum relations.
To do this, we observe that the energy $E(L)$ and velocity $\vel(L)$ depend parametrically on the radius of the vortex $L$.  Hence, we may construct the parametric dependence of the momentum $\mom(L)$ via the Hamiltonian equation
\begin{equation}
\vel(L)=\frac{\partial}{\partial p}E(p)\Bigg\vert_{p=P(L)}
=\frac{\partial E/\partial L}{\partial\mom/\partial L},
\label{eq:hamilton-equation}
\end{equation}
in view of which we have
\begin{equation}
\frac{\partial\mom}{\partial L}=
\frac{\partial E/\partial L}{\vel(L)}
\approx 2 \pi\,\Dsl\, \surf_{D-1}\, L^{\Dsl},
\end{equation}
where we have retained only the leading behavior at large $L$.
By integrating with respect to $L$ we obtain the parametric dependence of the momentum on the vortex radius,
\begin{equation}
\mom(L)\approx 2\pi \frac{D-2}{D-1} \surf_{D-1}\, L^{D-1},
\label{eq:partial-momentum}
\end{equation}
and by eliminating $L$ in favour of $E$ we arrive at the energy-momentum relation
\begin{equation}
E\approx \eta(D)\, \mom^{\frac{D-2}{D-1}} \left( \ln\frac{\mom \vcs}{\hbar}
+ \ln\frac{M}{\mdc \vcs^D} \right),
\end{equation}
where $\eta(D)$ is a $D$-dependent constant given by
\begin{equation}
\eta(D)=\frac{(\pi \surf_{D-1})^{1/(D-1)}}{D-1}
\left(\frac{D-1}{2D-4}\right)^{\frac{D-2}{D-1}},
\end{equation}
and the sub-leading logarithmic term is sensitive to the number of particles per coherence volume.

\section{%
\label{ScalingDim}
Scaling and dimensional analysis}
To ease the presentation, we have chosen to drop factors of $\hbar$, $M$ and $\mdc$ from our derivations, restoring physical units only when giving results at the ends of the various sections.  It is worth noting, however, that many of our results can be obtained via scaling and dimensional analysis, up to overall dimension-dependent factors and cutoff-dependent logarithms, as we now discuss.

There are two characteristic length-scales: the linear dimension of a vortex $L$, and the vortex core size $\vcs$.  The latter serves as a cutoff, and therefore only features in logarithmic factors, which are inaccessible to our scaling and dimensional analysis.  There are two characteristic mass-scales: the mass of each of the condensing particles $M$, and $\mdc L^D$, where $\mdc$ is the mass-density of the superfluid.  Lastly, there is a characteristic frequency-scale: $\hbar/ML^{2}$.

To ascertain the scaling form of the velocity of a vortex, we note that its derivation involves only the generalized $D$-dimensional Amp\`ere-Maxwell law, and thus is insensitive to $\mdc$.  Hence, dimensional analysis suggests that
\begin{equation}
\vel\sim
L\times
\frac{\hbar}{ML^2}
\sim
\frac{\hbar}{ML},
\end{equation}
which agrees, e.g., with Eq.~(\ref{eq:vel-gen-D}), except for numerical and logarithmic factors.
As for the energy of a flow associated with a vortex, it is proportional to the density $\mdc$, and therefore dimensional analysis suggests that
\begin{equation}
E\sim
\mdc L^{D}\times
L^2 \times
\left(\frac{\hbar}{M L^{2}}\right)^{2}
\sim
\mdc L^{\Dsl}\frac{\hbar^2}{M^2},
\end{equation}
which agrees, e.g., with Eq.~(\ref{eq:energy-asy-anyD}), except for numerical and logarithmic factors.
Regarding the momentum, it was computed from the Hamiltonian equation of motion~(\ref{eq:hamilton-equation}), and thus, like the energy, scales as $\mdc^{1}$.  Dimensional analysis therefore suggests that
\begin{equation}
\mom
\sim
\mdc L^D\times
L \times
\frac{\hbar}{M L^2}
\sim
\mdc L^{D-1}\frac{\hbar}{M},
\end{equation}
which agrees, e.g., with Eq.~(\ref{eq:partial-momentum}), except for numerical and logarithmic factors.
Concerning the frequency of oscillations of distortions of wave-vector $q$ of a hyper-planar vortex [see Eq.~(\ref{eq:en-osc-modes})], these are driven by the flow velocity, and therefore do not depend on $\mdc$.  Moreover, the undistorted vortex does not have a system-size-independent characteristic size; instead, the characteristic length-scale is set by $q^{-1}$.  These scaling notions, together with dimensional analysis, suggest that
\begin{equation}
\omega
\sim
\frac{\hbar}{M}\,q^2,
\end{equation}
consistent with Eq.~(\ref{eq:en-osc-modes}), except for numerical and logarithmic factors.

\section{%
\label{sec:conclusions}
Concluding remarks}
We have explored the extension, to arbitrary spatial dimension, of Onsager-Feynman quantized vortices in the flow of superfluid helium-four, focusing on the structure and energetics of the associated superflow and the corresponding dynamics of the vortices. To do this, we have analyzed the superflow that surrounds the vortices at equilibrium by invoking an extension, to arbitrary dimensions, of the three-dimensional analogy between: (i)~the magnetic field of Amp\`ere-Maxwell magnetostatics in the presence of specified, conserved electric currents concentrated on infinitesimally thin loops; and (ii)~the velocity field of equilibrium superfluidity of helium-four in the presence of specified, conserved vorticity concentrated on infinitesimally thin vortices.

By constructing the appropriate conditions that the velocity field must obey at equilibrium, if the flow is to have a given vorticial content, we have determined the corresponding flow in terms of this vorticial content, via the introduction of a suitable, skew-symmetric gauge potential of rank two fewer than the spatial dimension.  The structure of the flows associated with generically shaped vortices has enabled us to develop results for the dynamics that vortices inherit from these flows in all dimensions, and for the way in which this dynamics reflects both the intrinsic and extrinsic geometry of the vortices when the vortices are large and smooth, relative to the vortex core size.  This structure has also allowed us to ascertain the velocities with which the higher-dimensional extensions of circular vortex rings (i.e., vortex spheres and hyper-spheres) propagate, as well as to analyze the higher-dimensional extension (to planes and hyper-planes) of the Kelvin problem of the vibrations of a line vortex.

For all dimensions, we have expressed the equilibrium kinetic energy of superflow in the presence of a generic vortex in terms of the shape of the vortex, and have identified the leading contribution to this energy---for vortices that are smooth and large, relative to the vortex core size---as being proportional to the volume of the sub-manifold supporting the vortex.  We have also computed the kinetic energy for the higher-dimensional extensions of vortex rings, and used these energies to determine the momenta, and energy-momentum relations, of these maximally symmetrical vortices.  In addition, we have shown how the structure of our results, if not their details, can be obtained via elementary arguments, including scaling analyses.

The geometry of the vortices---both the intrinsic geometry of their shapes and the extrinsic geometry of the manner in which the vortices are embedded in the high-dimension ambient space---plays a central role in our developments.  The natural language for such developments is that of exterior calculus and differential forms, but we have not made this language essential, using instead the equivalent, but more widely known, language of skew-symmetric tensor fields.

Further issues that one might consider within the present framework include lattices of higher-dimensional extensions of vortices associated with equilibrium states of a rotating superfluid, along with the collective vibrations of such lattices (i.e., Tkachenko modes), as well as Magnus-type forces on vortices provided by background flows.  One might also consider the energetics and motion of knotted vortices in three dimensions (cf.~Kelvin's \lq\lq vortex atoms\rq\rq~\cite{ref:Kelvin-atom}) and---where mathematically available---topologically more exotic, higher-dimensional vortices, including higher-genus surfaces.  Moreover, one might apply the the energetics of higher-dimensional extensions of vortices, and the interactions between them, to a re-examination of the statistical mechanics of the superfluid-to-normal phase transition (i.e., the \lq\lq lambda\rq\rq\ transition) in higher dimensions.  Still more challenging would be the generalization of the circle of ideas discussed in this paper to the non-linear setting provided by non-Abelian gauge fields.

\ack
This work was supported by DOE Grant No.~DEFG02-91ER45439 through the Frederick Seitz Materials Research Laboratory at the University of Illinois.
It arose as an outgrowth of conversations with Gary Williams, whose encouragement we gratefully acknowledge.
We also gratefully acknowledge numerous informative conversations with John D'Angelo, Sheldon Katz and Michael Stone.
PMG thanks for its hospitality the Aspen Center for Physics, where part of this work was done.

\appendix

\section{%
\label{ap:derivation-stokes-D}
Source terms and quantized circulation}
In this appendix, we discuss two properties of the source field $\source$.  First, we show that the circulation, Eq.~(\ref{eq:ampere-int}), of the superflow around the vortex sub-manifold associated with the source term given in Eq.~(\ref{eq:current-allD}) is quantized to unit value.  We then show that the source field is divergenceless.

\begin{figure}[htbp]
    \centering
	\includegraphics[width=.40\textwidth]{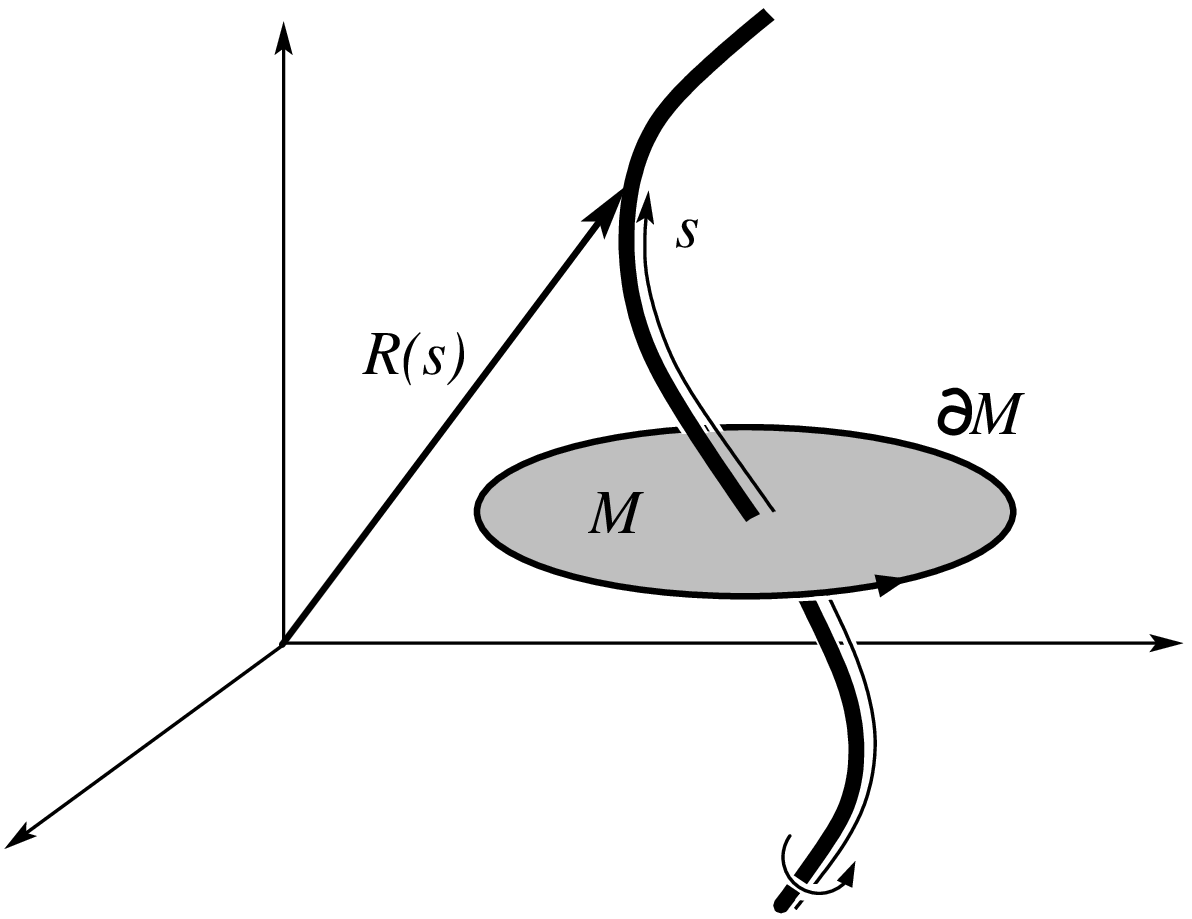}
	\caption{A closed path $\contourclosed$ around a segment of a vortex loop specified by $\dfc(\sig)$,
    with arc-length parameter $\sig$.  $\contourclosed$ is the boundary of the two-dimensional region
    $\contourinside$, which is punctured by the $\Dsl$-dimensional vortex.}
\label{FIG:ampere-law}
\end{figure}
We begin by considering a closed, one-dimensional path $\contourclosed$, centered on the vortex sub-manifold and encircling it, along with the associated two-dimensional region $\contourinside$ bounded by $\contourclosed$ and is punctured by the vortex sub-manifold, as shown in Fig.~\ref{FIG:ampere-law}.  In the language of differential forms, Stokes' theorem, applied to Eq.~(\ref{eq:ampere-DF-dual}), determines that the circulation $\vsk$ around $\contourclosed$ is given by
\begin{eqnarray}
\vsk
&=&
\oint_{\contourclosed}\vel_{d}(\pos)\,d\pos_{d}
=
\oint_{\contourclosed} \vel
\\
&=&
\int_{\contourinside}d\vel
=2\pi\int_{\contourinside}\sourcedual
=\frac{2\pi}{2!}\int_{\contourinside}
(\sourcedual)_{d_{1}d_{2}}(\pos)\,
d\pos_{d_{1}} \wedge d\pos_{d_{2}}\,.
\label{eq:int-source}
\end{eqnarray}
We take $\contourclosed$ to be circular, with the points on it being given by
\begin{equation}
\pos=\dfc(\sig_{0})+a_{1}N^{1}+a_{2}N^{2},
\label{eq:circle-area}
\end{equation}
where $a_{1}$ and $a_{2}$ vary subject to the constraint
$a_{1}^{2}+a_{2}^{2}={\rm const.}>0$,
and $N^{1}$ and $N^{2}$ are any pair of mutually perpendicular unit vectors that are perpendicular to the vortex sub-manifold at the point $\sig_{0}$.  An elementary computation then shows that
$d\pos_{d_{1}} \wedge d\pos_{d_{2}}
=da_{1}\wedge da_{2}\,(N^{1}\wedge N^{2})_{d_{1}d_{2}}$,
so that, from Eq.~(\ref{eq:int-source}), the circulation becomes
\begin{eqnarray}
\vsk=
\frac{2\pi}{2!}\int_{\contourinside}
(\sourcedual)_{d_{1}d_{2}}(\pos)\,
(N^{1}\wedge N^{2})_{d_{1}d_{2}}\,
da_{1}\wedge da_{2}.
\end{eqnarray}
Next, we insert the explicit components $\sourcedual_{d_{1}d_{2}}$, Eq.~(\ref{eq:current-allD-dual}), for the specific form of $\contourinside$ defined by Eq.~(\ref{eq:circle-area}), to arrive at
\begin{eqnarray}
\vsk&=&
\frac{2\pi}{2!}\int_{\contourinside}
\int d^{\Dsl}\sig\,
\epsilon_{d_{1}{\cdots}d_{\Dsl}d_{D-1}d_{D}}
(\newpart_{\inin_{1}}\dfc_{d_{1}})
\cdots
(\newpart_{\inin_{\Dsl}}\dfc_{d_{\Dsl}})\,
(N^{1}\wedge N^{2})_{d_{D-1}d_{D}}
\nonumber\\
&&\qquad\quad\times
\dirac^{(D)}\left(\dfc(\sig_{0})+a_{1}N^{1}+a_{2}N^{2}-\dfc(\sig)\right)\,
da_{1}\wedge da_{2}.
\end{eqnarray}
Next, we Taylor-expand $\dfc(\sig)$ in the argument of the delta function about the point $\sig=\sig_{0}$ and use the orthogonality of $N^{1}$ and $N^{2}$ with the tangent vectors $\{\partial\dfc/\partial\sig_{\inin}\}_{\inin=1}^{\Dsl}$ at $\sig_{0}$ to obtain
\begin{eqnarray}
\vsk&=&
\frac{2\pi}{2!}
\epsilon_{d_{1}{\cdots}d_{\Dsl}d_{D-1}d_{D}}
(\newpart_{\inin_{1}}\dfc_{d_{1}})
\cdots
(\newpart_{\inin_{\Dsl}}\dfc_{d_{\Dsl}})\,
(N^{1}\wedge N^{2})_{d_{D-1}d_{D}}\Big\vert_{\sig=\sig_{0}}
\nonumber\\&&
\qquad\quad\times
\int d^{\Dsl}\sig\,
\dirac^{(D-2)}\Big((\sig-\sig_{0})_{\inin}\,\partial_{\inin}\dfc(\sig)\vert_{\sig=\sig_{0}}\Big)
\nonumber\\&&
\qquad\qquad\qquad\times
\int\delta^{(1)}(a_{1})\,da_{1}\,
\int\delta^{(1)}(a_{2})\,da_{2}\,.
\end{eqnarray}
The last two integrals each give a factor of unity.  The remaining one, over $\sig$, is readily seen to give a factor of $1/\vert\det(\partial_{\inin}\dfc)\vert$ provided we invoke the fact that the term $\partial_{\inin}\dfc$ is a square $\Dsl\times\Dsl$ matrix in the subspace of vectors tangent to the vortex sub-manifold.  As we discuss
in Section~\ref{sec:ampere}, the factor
$\epsilon_{d_{1}{\cdots}d_{\Dsl}d_{D-1}d_{D}}
(\newpart_{\inin_{1}}\dfc_{d_{1}})\cdots
(\newpart_{\inin_{\Dsl}}\dfc_{d_{\Dsl}})$
is given by $(N^{1}\wedge N^{2})_{d_{D-1}d_{D}}\,\det(\partial_{\inin}\dfc)$.  Hence, in the formula for the circulation the determinants cancel, and the remaining two factors of the form $N^{1}\wedge N^{2}$ contract with one another to give a factor of 2, so that we arrive at the result
\begin{equation}
\vsk=\pm 1,
\end{equation}
the sign depending on the sense in which the encircling path $\contourclosed$ is followed.  Thus, we see that the
circulation of the superflow around the vortex sub-manifold associated with the source term given in Eq.~(\ref{eq:current-allD}) is indeed quantized to unit value.

In order to discuss the divergencelessness property of $\source$, we should distinguish between two cases.  In the case of $D=3$ the source field is a vector, which consequently has no skew-symmetric properties.  In arbitrary dimension $D$ greater than or equal to $4$, the source field $\source$ has a tensorial structure that we exploit in our analysis.  Let us consider first the case case of $D=3$.  The source term is then given by
\begin{equation}
\source_{d_1}(\pos)=\int d \sig \frac{d\dfc_{d_{1}}(\sig)}{d \sig}
\int \frac{d^3 q}{(2\pi)^3} e^{-i q\cdot(\pos-\dfc(\sig))},
\end{equation}
and its divergence may be manipulated as follows:
\begin{eqnarray}
\nabla_{d_{1}}\source_{d_{1}}(\pos)&=&\int d \sig \frac{d\dfc_{d_{1}}(\sig)}{d \sig} \int \frac{d^3q}{(2\pi)^3} (-iq_{d_{1}}) e^{-i q\cdot(\pos-\dfc(\sig))}
\nonumber
\\
&=&\int d \sig \int \frac{d^3q}{(2\pi)^3} (-iq_{d_{1}}) \frac{d\dfc_{d_{1}}(\sig)}{d \sig}  e^{-iq\cdot(\pos-\dfc(\sig))}
\nonumber
\\
&=&\int \frac{d^3q}{(2\pi)^3} \int d \sig\,  \frac{d}{d \sig} \left(e^{-i q\cdot(\pos-\dfc(\sig))}\right).
\label{eq:divJ-D3}
\end{eqnarray}
The integrand in the last line of Eq.~(\ref{eq:divJ-D3}) is a total derivative of a periodic function of $\sig$, and therefore the integral vanishes.

For $D\ge 4$ the source term has the structure
\begin{equation}
\source_{d_{1}\cdots d_{\Dsl}}(\pos)
=\!\!\int\!\!
d^{\Dsl}\sig\,
\varepsilon_{\inin_{1}\cdots\inin_{\Dsl}}
(\newpart_{\inin_{1}}\dfc_{d_{1}})
\cdots
(\newpart_{\inin_{\Dsl}}\dfc_{d_{\Dsl}})\,
\!\!\int\!\!\frac{d^D q}{(2\pi)^D}\,e^{-i q\cdot(\pos-\dfc(\sig))},
\end{equation}
and its divergence may be manipulated as follows:
\begin{eqnarray}
&&\nabla_{d_{1}}\source_{d_{1}\cdots d_{\Dsl}}(\pos)
=\int d^{\Dsl}\sig\,
\varepsilon_{\inin_{1}\cdots\inin_{\Dsl}}
(\newpart_{\inin_{1}}\dfc_{d_{1}})
\cdots
(\newpart_{\inin_{\Dsl}}\dfc_{d_{\Dsl}})
\nonumber \\
&&\qquad \qquad \qquad \qquad \times
\int \frac{d^{D}q}{(2\pi)^{D}} (-iq_{d_{1}}) e^{-i q\cdot(\pos-\dfc(\sig))}
\nonumber
\\
&&\qquad=
\int d^{\Dsl}\sig\int \frac{d^{D}q}{(2\pi)^{D}}\,
\varepsilon_{\inin_{1}\cdots\inin_{\Dsl}}
\big[
\newpart_{\inin_{1}}\big(e^{-iq\cdot(\pos-\dfc(\sig))}\big)
\big]
(\newpart_{\inin_{2}}\dfc_{d_{2}})
\cdots
(\newpart_{\inin_{\Dsl}}\dfc_{d_{\Dsl}})
\nonumber
\\
&&\qquad=
\int \frac{d^{D}q}{(2\pi)^{D}} \int d^{\Dsl}\sig
\Big\{
\varepsilon_{\inin_{1}\cdots\inin_{\Dsl}}
  \newpart_{\inin_{1}}
\big[
e^{-iq\cdot(\pos-\dfc(\sig))}
(\newpart_{\inin_{2}}\dfc_{d_{2}})
\cdots
(\newpart_{\inin_{\Dsl}}\dfc_{d_{\Dsl}})
\big]
\nonumber
\\
&&\qquad\qquad
 -e^{-iq\cdot(\pos-\dfc(\sig))}
\varepsilon_{\inin_{1}\cdots\inin_{\Dsl}}
\newpart_{\inin_{1}}
\big[
(\newpart_{\inin_{2}}\dfc_{d_{2}})
\cdots
(\newpart_{\inin_{\Dsl}}\dfc_{d_{\Dsl}})
\big]
\Big\}.
\label{eq:divJ-D}
\end{eqnarray}
In the last line of Eq.~(\ref{eq:divJ-D}) the first term vanishes because the integrand is a total derivative of a periodic function, whilst the second term vanishes via the skew-symmetric properties of Levi-Civita symbol.  Hence, we have the divergencelessness condition $\nabla_{d_{1}}\source_{d_{1}\cdots d_{\Dsl}}=0$.

\section{%
\label{ap:velocity-int}
Computing the velocity field}
As shown in Section~\ref{sec:fixeqm}, the velocity field due to a single, $\Dsl$-dimensional vortex is given by
\begin{eqnarray}
\vel_{d_{D}}(x)
&=&
-\frac{2\pi}{\Dsl\,!}\,
\epsilon_{d_{1}\cdots d_{D}}
\int\dbar^{D}q\,
\frac{iq_{d_{D-1}}}{q\cdot q}
{\rm e}^{-iq\cdot\pos}
\nonumber
\\
&&\qquad\qquad\times
\int d^{\Dsl}\sig\,
\varepsilon_{\inin_{1}\cdots\inin_{D-2}}
(\newpart_{\inin_{1}}\dfc_{d_{1}})
\cdots
(\newpart_{\inin_{\Dsl}}\dfc_{d_{\Dsl}})\,
{\rm e}^{iq\cdot\dfc(\sig)}.
\end{eqnarray}
Performing the $q$ integration using Eq.~(\ref{eq:Coulomb-int-one}) of \ref{ap:invariant-int}, and noting the formula for the surface area of a $D$-dimensional sphere of unit radius~\cite{ref:levyleblond},
the velocity field becomes
\begin{eqnarray}
\vel_{d_{D}}(x)
&=&
-\frac{2\pi}{\Dsl\,!}\,
\epsilon_{d_{1}\cdots d_{D}}
\int d^{\Dsl}\sig\,
\varepsilon_{\inin_{1}\cdots\inin_{\Dsl}}
(\newpart_{\inin_{1}}\dfc_{d_{1}})
\cdots
(\newpart_{\inin_{\Dsl}}\dfc_{d_{\Dsl}})
\nonumber
\\
&&\qquad\qquad\times
\int\dbar^{D}q\,
\frac{iq_{d_{D-1}}}{q\cdot q}
{\rm e}^{iq\cdot(\dfc(\sig)-\pos)}
\nonumber
\\
&&=
\frac{2\pi}{\Dsl\,!}\,
\epsilon_{d_{1}\cdots d_{D}}
\int d^{\Dsl}\sig\,
\varepsilon_{\inin_{1}\cdots\inin_{\Dsl}}
(\newpart_{\inin_{1}}\dfc_{d_{1}})
\cdots
(\newpart_{\inin_{\Dsl}}\dfc_{d_{\Dsl}})
\nonumber
\\
&&\qquad\qquad\times
\partial_{d_{D-1}}
\int\dbar^{D}q\,
\frac{{\rm e}^{iq\cdot(\dfc(\sig)-\pos)}}{q\cdot q}
\nonumber
\\
&&=
\frac{\pi^{\frac{3}{2}-D}}{\Dsl\,!\,4}\surf_{D-1}\,
\Gamma\big(\textstyle{\frac{D}{2}-\frac{1}{2}}\big)\,
\Gamma\big(\textstyle{\frac{\Dsl}{2}}\big)\,
\epsilon_{d_{1}\cdots d_{D}}
\nonumber
\\
&&\qquad\times
\int d^{\Dsl}\sig\,
\varepsilon_{\inin_{1}\cdots\inin_{\Dsl}}
(\newpart_{\inin_{1}}\dfc_{d_{1}})
\cdots
(\newpart_{\inin_{\Dsl}}\dfc_{d_{\Dsl}})\,
\left(\partial_{d_{D-1}}\big\vert{\pos-\dfc(\sig)}\big\vert^{-\Dsl}\right)
\nonumber
\\
&&=
\frac{\pi^{\frac{3}{2}-D}}{(\Dsl-1)!\,4}\surf_{D-1}\,
\Gamma\big(\textstyle{\frac{D}{2}-\frac{1}{2}}\big)\,
\Gamma\big(\textstyle{\frac{\Dsl}{2}}\big)\,
\epsilon_{d_{1}\cdots d_{D}}
\nonumber
\\
&&\qquad\times
\int d^{\Dsl}\sig\,
\varepsilon_{\inin_{1}\cdots\inin_{\Dsl}}
(\newpart_{\inin_{1}}\dfc_{d_{1}})
\cdots
(\newpart_{\inin_{\Dsl}}\dfc_{d_{\Dsl}})\,
\frac{\big(\pos-\dfc(\sig)\big)_{d_{D-1}}}{\!\!\!\!\!\!\big\vert{\pos-\dfc(\sig)}\big\vert^{D}}
\nonumber
\\
&&=
\frac{1}{2\,\pi^{\Dsl/2}}\,\frac{\Gamma(\Dsl/2)}{(\Dsl-1)!}\,
 \epsilon_{d_{1}\cdots d_{D}}
\nonumber
\\
&&\qquad\times
\int d^{\Dsl}\sig\,
\varepsilon_{\inin_{1}\cdots\inin_{\Dsl}}
(\newpart_{\inin_{1}}\dfc_{d_{1}})
\cdots
(\newpart_{\inin_{\Dsl}}\dfc_{d_{\Dsl}})\,
\frac{\big(\pos-\dfc(\sig)\big)_{d_{D-1}}}{\!\!\!\!\!\!\big\vert{\pos-\dfc(\sig)}\big\vert^{D}}.
\label{eq-ap:vel-formula-surf}
\end{eqnarray}

\section{%
\label{ap:invariant-int}
Invariant integrals}
Consider the $q$ integral in Eq.~(\ref{eq:q-int-for-energy}).  Writing $\sep$ for
$\left(\dfc(\sig^{\prime})-\dfc(\sig)\right)$, and invoking
rotational symmetry and dimensional analysis, we have
\begin{eqnarray}
&&(2\pi)^{D}\int\dbar^{D}q\,
{\rm e}^{iq\cdot\sep}
\frac{q_{d^{\prime}}\,q_{d}}{(q\cdot q)^{2}}
=\frac{1}{\left(\sep\cdot\sep\right)^{\Dsl/2}}
\left(
\delta_{d^{\prime}d}\,{\cal P}_{1}+
\frac{\sep_{d^{\prime}}\,\sep_{d}}{\sep\cdot\sep}\,{\cal P}_{2}
\right).
\end{eqnarray}
Contracting first with $\delta$ and then $\sep_{d^{\prime}}\,\sep_{d}$,
we arrive at the invariant integrals
\begin{eqnarray}
\phantom{\frac{1}{(\sep\cdot\sep)}}
\int d^{D}q\,
{\rm e}^{iq\cdot\sep}
\frac{1}{q\cdot q}
&=&
\frac
{D{\cal P}_{1}+{\cal P}_{2}}
{\left(\sep\cdot\sep\right)^{\Dsl/2}},
\label{eq:Coulomb-int-one}
\\
\frac{1}{(\sep\cdot\sep)}
\int d^{D}q\,
{\rm e}^{iq\cdot\sep}
\frac{(q\cdot\sep)^{2}}{(q\cdot q)^{2}}
&=&
\frac{{\cal P}_{1}+{\cal P}_{2}}
{\left(\sep\cdot\sep\right)^{\Dsl/2}}.
\label{eq:Coulomb-int-two}
\end{eqnarray}
The integrals~(\ref{eq:Coulomb-int-one}) and (\ref{eq:Coulomb-int-one}) can be computed in hyper-spherical coordinates, and we thus obtain for ${\cal P}_{1}$ and ${\cal P}_{2}$:
\begin{eqnarray}
{\cal P}_{1}&=&
\phantom{(2-D)}\,
\surf_{D-1}\,\sqrt{\pi}\,\Gamma\left(\textstyle{\frac{D}{2}-\frac{1}{2}}\right)\,
2^{D-4}\,\Gamma\left(\textstyle{\frac{D}{2}-1}\right),
\\
{\cal P}_{2}&=&
(2-D)\,
\surf_{D-1}\,\sqrt{\pi}\,\Gamma\left(\textstyle{\frac{D}{2}-\frac{1}{2}}\right)\,
2^{D-4}\,\Gamma\left(\textstyle{\frac{D}{2}-1}\right),
\end{eqnarray}
where $\surf_{D}$ is the surface area of a $D$-dimensional sphere of unit radius~\cite{ref:levyleblond}, and $\Gamma(z)$ is the standard Gamma function~\cite{ref:Abram-handbook}.

\section{%
\label{ap:vortex-vel}
Asymptotic analysis for the velocity of a vortex}
To determine the asymptotic behaviour of the velocity, we start with Eq.~(\ref{eq:vel-local}), which contains the factor
\begin{equation}
\int d^{\Dsl}\sig\,
\varepsilon_{\inin_{1}\cdots\inin_{\Dsl}}
\newpart_{\inin_{1}}\dfc_{d_{1}}(\sig)
\cdots
\newpart_{\inin_{\Dsl}}\dfc_{d_{\Dsl}}(\sig)\,
\frac{\big(\dfc(\sigo)-\dfc(\sig)\big)_{d_{D-1}}}
{\big\vert{\dfc(\sigo)-\dfc(\sig)}\big\vert^{D}},
\label{eq:vel-expand}
\end{equation}
and Taylor-expand the integrand about the point $\sigo$ to obtain
\begin{eqnarray}
&&
\phantom{\Big\{}
\int d^{\Dsl}\sig\,
\bigg\{
\Big[(\newpart_{\inin_{1}}\dfc_{d_{1}})\cdots
(\newpart_{\inin_{\Dsl}}\dfc_{d_{\Dsl}})\big\vert_{\sig=\sigo}
\nonumber\\
    &&\qquad\qquad+
(\newpart_{\inin_{1}}\dfc_{d_{1}})\cdots
(\newpart_{\nd}\newpart_{\inin_{\md}}\dfc_{d_{\md}})\cdots
(\newpart_{\inin_{\Dsl}}\dfc_{d_{\Dsl}})\big\vert_{\sig=\sigo}(\sig_{\nd}-\sigo_{\nd})
+
\cdots\Big]
\nonumber\\
    &&\quad\,\phantom{\Big\{}\,\,\,\,
\times 
\Big[-
(\newpart_{\nd}\dfc_{d_{D-1}})
\big\vert_{\sig=\sigo}(\sig_{\nd}-\sigo_{\nd})
\nonumber\\
    &&\qquad\qquad
-\frac{1}{2}
(\newpart_{\nd}\newpart_{\ndp}\dfc_{d_{D-1}})
\big\vert_{\sig=\sigo}(\sig_{\nd}-\sigo_{\nd})(\sig_{\ndp}-\sigo_{\nd})
+\cdots\Big]
\nonumber\\
    &&\quad\,\,\,\,\,
\times 
\Big(
\Big[
-(\newpart_{\nd}\dfc_{\deflambda})
\big\vert_{\sig=\sigo}
(\sig_{\nd}-\sigo_{\nd})
\nonumber\\
    &&\qquad\qquad
-\frac{1}{2}
(\newpart_{\nd}\newpart_{\ndp}\dfc_{\deflambda})
\big\vert_{\sig=\sigo}
(\sig_{\nd}-\sigo_{\nd})(\sig_{\ndp}-\sigo_{\ndp})
+\cdots\Big]
\nonumber\\
    &&\quad\,\phantom{\Big\{}\,\,\,
\times 
\Big[
-(\newpart_{\nd}\dfc_{\deflambda})
\big\vert_{\sig=\sigo}
(\sig_{\nd}-\sigo_{\nd})
\nonumber\\
    &&\qquad\qquad
-\frac{1}{2}
(\newpart_{\nd}\newpart_{\ndp}\dfc_{\deflambda})
\big\vert_{\sig=\sigo}
(\sig_{\nd}-\sigo_{\nd})(\sig_{\ndp}-\sigo_{\ndp})
+\cdots\Big]\Big)^{-\frac{D}{2}}
\label{eq:vel-keep}
\end{eqnarray}
Next, we expand the terms in square brackets, retaining only terms to leading order, and collect terms of similar structure in the integration variables $\sig$, observing that two such structures arise:
\begin{eqnarray}
{\cal I}^{(2)}_{\inin_{1}\inin_{2}}(\sigo)
&:=&
\int d^{\Dsl}\!\sig\,
\frac{
(\sig_{\inin_{1}}-\sigo_{\inin_{1}})
(\sig_{\inin_{2}}-\sigo_{\inin_{2}})}
{\left[
(\sig_{\inin}-\sigo_{\inin})\,
\met_{\inin\inin^{\prime}}(\sigo)\,
(\sig_{\inin^{\prime}}-\sigo_{\inin^{\prime}}
)\right]^{D/2}},
\\
\noalign{\medskip}
{\cal I}^{(4)}_{\inin_{1}\inin_{2}\inin_{3}\inin_{4}}
&:=&
\int d^{\Dsl}\!\sig\,
\frac{
(\sig_{\inin_{1}}-\sigo_{\inin_{1}})(\sig_{\inin_{2}}-\sigo_{\inin_{2}})
(\sig_{\inin_{3}}-\sigo_{\inin_{3}})(\sig_{\inin_{4}}-\sigo_{\inin_{4}})
      }
{
\left[
(\sig_{\inin}-\sigo_{\inin})\,
\met_{\inin\inin^{\prime}}(\sigo)\,
(\sig_{\inin^{\prime}}-\sigo_{\inin^{\prime}})
\right]^{(D/2)+1}
}.
\end{eqnarray}
The asymptotic behaviours of these structures are given by
\begin{eqnarray}
{\cal I}^{(2)}_{\inin_{1}\inin_{2}}(\sigo)
&\approx&
\frac{\surf_{\Dsl}}{\Dsl}\,
\frac{\metinv_{\inin_{1}\inin_{2}}}{\sqrt{{\rm det}\,\met}}\,
\ln(L/\vcs),
\\
\noalign{\medskip}
{\cal I}^{(4)}_{\inin_{1}\inin_{2}\inin_{3}\inin_{4}}(\sigo)
&\approx&
\frac{\surf_{\Dsl}}{D\Dsl}\,
\frac{
\metinv_{\inin_{1}\inin_{2}}\,\metinv_{\inin_{3}\inin_{4}}+
\metinv_{\inin_{1}\inin_{3}}\,\metinv_{\inin_{2}\inin_{4}}+
\metinv_{\inin_{1}\inin_{4}}\,\metinv_{\inin_{2}\inin_{3}}
}{\sqrt{{\rm det}\,\met}}\,
\ln(L/\vcs),
\end{eqnarray}
where $\metinv:=\met^{-1}$, and ${\text\met}$ and ${\text\metinv}$ are evaluated at $\sigo$.

\section{%
\label{ap:shape-operators}
Geometric content of the vortex velocity}
Equation~(\ref{eq:vel-local-approx}) for the asymptotic approximation to the vortex velocity $\velv_{d_{D}}(\sigo)$ contains three terms,
\begin{eqnarray}
\term^{1}_{d_{D}}&:=&
\epsilon_{d_{1}\cdots d_{D}}
\epsilon_{\inin_{1}\cdots\inin_{\Dsl}}
(\newpart_{\inin_{1}}\dfc_{d_{1}})
\cdots
(\newpart_{\inin_{\Dsl}}\dfc_{d_{\Dsl}})
(\newpart_{\nd}\newpart_{\ndp}\dfc_{d_{D-1}})
\metinv_{\nd\ndp}\,,
\label{eq:app-int-one}
\\
\term^{2}_{d_{D}}&:=&
\epsilon_{d_{1}\cdots d_{D}}
\epsilon_{\inin_{1}\cdots\inin_{\Dsl}}
(\newpart_{\ndp}\dfc_{d_{D-1}})
(\newpart_{\inin_{1}}\dfc_{d_{1}})\cdots
    \nonumber\\&&\qquad\qquad\qquad
\times
(\newpart_{\nd}\newpart_{\inin_{\md}}\dfc_{d_{\md}})
\cdots
(\newpart_{\inin_{\Dsl}}\dfc_{d_{\Dsl}})
\metinv_{\nd\ndp}\,,
\\
\term^{3}_{d_{D}}&:=&
\epsilon_{d_{1}\cdots d_{D}}
\epsilon_{\inin_{1}\cdots\inin_{\Dsl}}
(\newpart_{\nd}\dfc_{\deflambda})
(\newpart_{\nnd}\newpart_{\nndp}\dfc_{\deflambda})
(\newpart_{\inin_{1}}\dfc_{d_{1}})
\cdots
(\newpart_{\inin_{\Dsl}}\dfc_{d_{\Dsl}})
(\newpart_{\ndp}\dfc_{d_{D-1}})
    \nonumber\\&&\qquad\qquad\qquad
\times
\big(\metinv_{\nd\nnd}\metinv_{\ndp\nndp}+
       \metinv_{\nd\nndp}\metinv_{\nnd\nndp}+
       \metinv_{\nd\ndp}\metinv_{\nnd\nndp}\big),
\end{eqnarray}
which we consider in turn.

The first term contains the ingredient
\begin{equation}
\epsilon_{d_{1}\cdots d_{D}}\,
\varepsilon_{\inin_{1}\cdots\inin_{\Dsl}}\,
(\newpart_{\inin_{1}}\dfc_{d_{1}})
\cdots
(\newpart_{\inin_{\Dsl}}\dfc_{d_{\Dsl}}).
\end{equation}
which is identical to the term in Eq.~(\ref{eq:geom-factor}), computed in Section~\ref{sec:ampere}.  By substituting its value into Eq.~(\ref{eq:app-int-one}), we obtain
\begin{equation}
\term^{1}_{d_{D}}=\Dsl!\,\sqrt{{\rm det}\met}\,\,
\shop_{d_{D-1}d_{D}}\,
(\newpart_{\nd}\newpart_{\ndp}\dfc_{d_{D-1}})\,
\metinv_{\nd\ndp}\,.
\end{equation}
It is straightforward to observe (via partial integration) that $\term^{2}$ is identical to $\term^{1}$, up to a sign.
As for $\term^{3}_{d_{D}}$, it is proportional to
\begin{equation}
\epsilon_{d_{1}\cdots d_{D}}\,
\varepsilon_{\inin_{1}\cdots\inin_{\Dsl}}
(\newpart_{\nd}\dfc_{\deflambda})
(\newpart_{\inin_{1}}\dfc_{d_{1}})
\!\cdots\!
(\newpart_{\inin_{\Dsl}}\dfc_{d_{\Dsl}})
(\newpart_{\ndp}\dfc_{d_{D-1}}),
\end{equation}
a term that is skew-symmetric in the indices $(d_1,\ldots,d_\Dsl)$ and symmetric under the interchange of $\deflambda$ and $d_i$.  With the only possible values for both $\deflambda$ and $d_i$ being in the set $\{1,\ldots,\Dsl\}$, $\term^{3}$ vanishes.

\section{%
\label{ap:exp-brane-vel}
Velocity for a weak distortion of a hyper-planar vortex}
In this appendix, we consider weak distortions of a $\Dsl$-dimensional hyper-planar vortex, and determine the velocity that the corresponding superflow confers on such distortions.
We begin with Eq.~(\ref{eq:nearlyflat}), which is Eq.~(\ref{eq:vel-local}) but with the replacement Eq.~(\ref{eq:height-function}), which yields the velocity $\velv(\sigoM)$ of points $\sigoM$ on the vortex in terms of $\height$.
We then expand $\velv$ in powers of $\height$, which we take to be small, and, for the sake of compactness, we temporarily omit an overall factor of $2\pi\Dsl{\Gamma\left(\Dsl/2\right)}{\big/}{4\pi^{D/2}}$
which multiplies the velocity.
Hence, we arrive at the formula
\begin{eqnarray}
\velv_{d_{D}}(\sigoM)
&=&
\epsilon_{d_{1}\cdots d_{D}}
\int d^{\Dsl}\sig\,
\newpart_{\inin_{1}}(\flat_{d_{1}}(\sig)+\height_{d_{1}}(\sig))
\cdots
\newpart_{\inin_{\Dsl}}(\flat_{d_{\Dsl}}(\sig)+\height_{d_{\Dsl}}(\sig))
\nonumber
\\
&&
\qquad\qquad
\times
\frac{\big(\flat(\sigoM)-\flat(\sig)+\height(\sigoM)-\height(\sig)\big)_{d_{D-1}}}
{\big\vert{\flat(\sigoM)-\flat(\sig)+\height(\sigoM)-\height(\sig)}\big\vert^{D}}
\\
\noalign{\medskip}
&=&
\epsilon_{d_{1}\cdots d_{D}}
\int d^{\Dsl}\sig\,
\newpart_{\inin_{1}}\flat_{d_{1}}(\sig)
\cdots
\newpart_{\inin_{\Dsl}}\flat_{d_{\Dsl}}(\sig)
\frac{\big(\flat(\sigoM)-\flat(\sig)\big)_{d_{D-1}}}
{\big\vert{\flat(\sigoM)-\flat(\sig)}\big\vert^{D}}
\nonumber
\\
&&
\quad
+
\epsilon_{d_{1}\cdots d_{D}} \sum_{\nu=_1}^{\Dsl}
\int d^{\Dsl}\sig\,
\newpart_{\inin_{1}}\flat_{d_{1}}(\sig)
\cdots
\newpart_{\inin_\nu}\height_{d_\nu} (\sig)
\cdots
\newpart_{\inin_{\Dsl}}\flat_{d_{\Dsl}}(\sig)
\nonumber
\\
\noalign{\smallskip}
&&
\qquad\qquad
\times
\frac{\big(\flat(\sigoM)-\flat(\sig)\big)_{d_{D-1}}}
{\big\vert{\flat(\sigoM)-\flat(\sig)}\big\vert^{D}}
\nonumber
\\
&&
\quad
+
\epsilon_{d_{1}\cdots d_{D}}
\int d^{\Dsl}\sig\,
\newpart_{\inin_{1}}\flat_{d_{1}}(\sig)
\cdots
\newpart_{\inin_{\Dsl}}\flat_{d_{\Dsl}}(\sig)
\frac{\big(\height(\sigoM)-\height(\sig)\big)_{d_{D-1}}}
{\big\vert{\flat(\sigoM)-\flat(\sig)}\big\vert^{D}}
\nonumber
\\
&&
\quad
-
D\epsilon_{d_{1}\cdots d_{D}}
\int d^{\Dsl}\sig\,
\newpart_{\inin_{1}}\flat_{d_{1}}(\sig)
\cdots
\newpart_{\inin_{\Dsl}}\flat_{d_{\Dsl}}(\sig)
\nonumber
\\
&&
\qquad\qquad
\times
\big(\flat(\sigoM)-\flat(\sig)\big)\cdot\big(\height(\sigoM)-\height(\sig)\big)
\frac{\big(\flat(\sigoM)-\flat(\sig)\big)_{d_{D-1}}}
{\big\vert{\flat(\sigoM)-\flat(\sig)}\big\vert^{D+2}}
\nonumber
\\
\noalign{\smallskip}
&&
\quad
+
{\cal{O}}\left(\height^2\right).
\label{eq:nearly-flat-expand}
\end{eqnarray}
As $\flat(\sig)$ describes a hyper-plane, its derivatives with respect to $\sig_{\inin}$ are simply Kronecker deltas: $\newpart_{\inin}\flat_{d}(\sig)=\delta_{\inin d}$.
Furthermore, the term of zeroth order in $\height$ on the right hand side of Eq.~(\ref{eq:nearly-flat-expand}), viz.,
\begin{eqnarray}
&&
\epsilon_{d_{1}\cdots d_{D}} \int d^{\Dsl}\sig\,
\delta_{\inin_{1}d_{1}} \cdots \delta_{\inin_{\Dsl}d_{\Dsl}}
\frac{\big(\sigoM-\sig\big)_{\inin_\nu} \delta_{\inin_\nu d_{D-1}}}
{\big\vert{\flat(\sigoM)-\flat(\sig)}\big\vert^{D}}
\nonumber
\\
&&\qquad\qquad\qquad=
\epsilon_{\inin_{1}\cdots \inin_{D-2} \inin_{\nu} d_{D}} \int d^{\Dsl}\sig\,
\frac{\big(\sigoM-\sig\big)_{\inin_\nu} }
{\big\vert{\flat(\sigoM)-\flat(\sig)}\big\vert^{D}},
\nonumber
\end{eqnarray}
vanishes because $\inin_\nu$ can only take values between $\inin_1$ and $\inin_{\Dsl}$ and the term is already skew-symmetric in these indices.  The last term of first order in $\height$ on the right hand side of Eq.~(\ref{eq:nearly-flat-expand}) also vanishes, via the orthogonality of $\flat$ and $\height$.

We observe that the remaining terms for the velocity, i.e.,
\begin{eqnarray}
&&\velv_{d_{D}}(\sigoM)
\approx
\sum_{\nu, \nu^{\prime}=1}^{\Dsl}\int d^{\Dsl}\sig\,\,
\epsilon_{12\cdots d_\nu \cdots D-2 \,d_{D-1} d_D}\,
\newpart_{\inin_\nu}\height_{d_\nu}(\sig)\,
\frac{\big(\sigoM-\sig\big)_{\nu^{\prime}}\,\delta_{\nu^{\prime}d_{D-1}}}
{\big\vert{\flat(\sigoM)-\flat(\sig)}\big\vert^{D}}
\nonumber
\\
&&\qquad\qquad\qquad
+\int d^{\Dsl}\sig\,
\epsilon_{12 \cdots D-2 \,d_{D-1} d_D}\,
\frac{\big(\height(\sigoM)-\height(\sig)\big)_{d_{D-1}}}
{\big\vert{\flat(\sigoM)-\flat(\sig)}\big\vert^{D}},
\end{eqnarray}
are nonzero only for the components $\velv_{D-1}$ and $\velv_{D}$.  In order to keep the notation compact, it is useful to introduce a new index, $\gamma$, which takes only the values $D-1$ and $D$.
As a last step, we rearrange the indices of the Levi-Civita symbols,
arriving at the $(D-1)^{\rm th}$ and $D^{\rm th}$ components of the velocity, to first order in the height:
\begin{eqnarray}
&&
\velv_{\gamma^{\prime}}(\sigoM)
\approx
\sum_{\gamma=D-1}^{D} \epsilon_{\gamma \gamma^{\prime}}
\int \frac{d^{\Dsl}\sig}{\big\vert{\flat(\sigoM)-\flat(\sig)}\big\vert^{D}}\,
\nonumber
\\
&&\qquad\qquad\qquad\qquad
\times
\Big(
\big(\height(\sigoM)-\height(\sig)\big)_{\gamma}
-\sum_{\nu=1}^{\Dsl} (\newpart_\nu \height_{\gamma}(\sig))(\sigoM-\sig)_\nu
\Big),
\end{eqnarray}
with all other components of $\velv$ vanishing to this order.

\section{%
\label{ap:four-an-brane}
Fourier transform of the velocity for the weak distortion of a hyper-planar vortex}
The required Fourier transform is of the convolution on the right-hand side of Eq.~(\ref{eq:vel-dominant}), and is of the form
\begin{eqnarray}
&&\int{d^{\Dsl}\sigoM}\,\,
{\rm e}^{-iq\cdot\sigoM}
\int d^{\Dsl}\sig\,
\frac{(\sigoM-\sig)_{\nu}(\sigoM-\sig)_{\bar{\nu}}}
{\left\vert{\sigoM-\sigo}\right\vert^{D}}\,
\frac{\partial^{2}\height_{\gamma}}
{\partial \sig_{\nu}\partial\sig_{\bar{\nu}}}
\\
&&=\phantom{-}(2\pi)^{\Dsl}
\int{d^{\Dsl}\sigoM}\,\,
{\rm e}^{-iq\cdot(\sigoM-\sig)}
\frac{(\sigoM-\sig)_{\nu}(\sigoM-\sig)_{\bar{\nu}}}
{\left\vert{\sigoM-\sigo}\right\vert^{D}}\,
\int{d^{\Dsl}\sig}\,
{\rm e}^{-iq\cdot\sig}
\frac{\partial^{2}\height_{\gamma}}
{\partial \sig_{\nu}\partial\sig_{\bar{\nu}}}
\\
&&=-
(2\pi)^{\Dsl}\,
\widehat{\height}_{\gamma}(q)\,
\int{d^{\Dsl}\sigoM}\,
{\rm e}^{-iq\cdot\sigoM}
\frac{(q\cdot\sigoM)^{2}}
{\left\vert{\sigoM}\right\vert^{D}}.
\end{eqnarray}
The remaining integral is readily computed in hyper-spherical coordinates:
\begin{eqnarray}
&&\int{d^{\Dsl}\sigoM}\,
{\rm e}^{-iq\cdot\sigoM}
\frac{(q\cdot\sigoM)^{2}}
{\left\vert{\sigoM}\right\vert^{D}}
\\
&&=
\vert{q}\vert^{2}
\surf_{\Dsl-1}\int
\sig^{\Dsl-1}\,ds\frac{1}{\sig^{D}}
\int_{0}^{\pi}d\theta\,
\sin^{\Dsl-2}\theta\,(\sig\cos\theta)^{2}
{\rm e}^{-iq\sig\cos\theta}
\\
&&=
\vert q\vert^2 \frac{\pi^{\Dsl/2}}{\Gamma(\Dsl/2)}\,\ln\left({1}/{\vert{q\vert}\xi}\right),
\end{eqnarray}
where we have imposed a short-distance cutoff at $\sig=\vcs$ to render the integral convergent.  The spatial Fourier transform evaluated in the present appendix, together with the temporal one, convert the differential (in time), integro-differential (in space) equation of motion~(\ref{eq:EOM-approx}) into algebraic form; see Eq.~(\ref{eq:EOM-algebraic}).


\begin{thebibliography}{100}
\bibitem{ref:levyleblond}
The term $\surf_{\Delta}$ is defined to be the area of the ($\deldef-1$)-dimensional
surface of a $\Delta$-dimensional hyper-sphere of unit radius, viz.,
$\surf_{\deldef}=2\Gamma(1/2)^{\deldef}\!\big/\Gamma(\deldef/2)$,
where $\Gamma(z)$ is the standard Gamma function~\cite{ref:Abram-handbook}.
Aspects of the geometry of hyper-spheres are discussed in Ref.~\cite{ref:levyleblond-art}.
\end{thebibliography}
\section*{References}


\end{document}